# Synergistic Niobium Doped Two-Dimensional Zirconium Diselenide: An Efficient Electrocatalyst for $O_2$ Reduction Reaction

Ashok Singh[1] and Srimanta Pakhira[1,2]*

[1] Theoretical Condensed Matter Physics and Advanced Computational Materials Science Laboratory, Department of Physics, Indian Institute of Technology Indore (IIT Indore), Simrol, Khandwa Road, Indore-453552, Madhya Pradesh, India.

[2] Theoretical Condensed Matter Physics and Advanced Computational Materials Science Laboratory, Centre for Advanced Electronics (CAE), Indian Institute of Technology Indore (IIT Indore), Simrol, Khandwa Road, Indore-453552, Madhya Pradesh, India.

*Corresponding author: spakhira@iiti.ac.in (or) spakhirafsu@gmail.com

**ABSTRACT**

The development of high-activity and low-price cathodic catalysts to facilitate the electrochemical sluggish $O_2$ reduction reaction (ORR) is very important to achieve the commercial application of fuel cells. Here, we have investigated the electrocatalytic activity of two-dimensional single-layer Nb-doped zirconium diselenide (2D Nb-$ZrSe_2$) towards ORR by employing the dispersion corrected Density Functional Theory (DFT-D) method. Through our study, we computed structural properties, electronic properties, and energetics of the 2D Nb-$ZrSe_2$ and ORR intermediates to analyze the electrocatalytic performance of the 2D Nb-$ZrSe_2$. The electronic properties calculations depict that the 2D monolayer $ZrSe_2$ has a large band gap of 1.48 eV, which is not favorable for the ORR mechanism. After the doping of Nb, the electronic band gap vanishes and 2D Nb-$ZrSe_2$ acts as a conductor. We studied both the dissociative and associative pathways through which the ORR can proceed to reduce the oxygen molecule ($O_2$). Our results show that the more favorable path for $O_2$ reduction on the surface of the 2D Nb-$ZrSe_2$ is the $4e^-$ associative path. The detailed ORR mechanisms (both associated and dissociative) have been explored by computing the changes of Gibbs free energy (ΔG). All the ORR reaction intermediate's steps are thermodynamically stable and energetically favorable. The free energy profile for the associative path shows the downhill behavior of the free energy vs. the reaction steps, suggesting that all ORR intermediate structures are catalytically active for the $4e^-$ associative path and a high $4e^-$ reduction pathway selectivity. Therefore, 2D Nb-ZrSe$_2$ is a promising catalyst for the ORR which can be used as an alternative ORR catalyst compared with expensive platinum (Pt).

**INTRODUCTION**

The escalating global population and the ongoing advancements in global industrialization and society have led to a significant surge in global energy demand. Nowadays a great part of the energy demand over the world is fulfilled by fossil fuels.[1] However, the bitter truth is that the traditional resources of fossil fuels are in danger of being depleted with indefinite extraction and utilization.[1] In addition to this, the consumption of fossil fuels results in the emission of pollutants. So, the concern about the energy crisis, environmental pollution, and climate change have put at risk the future of human society.[2] The gradual exhaustion of fossil fuels and the detrimental impact on the environment caused by pollution have provoked scientists and researchers over the world to develop renewable, sustainable, eco-friendly, and highly efficient energy resources.[3] Renewable energy systems including fuel cells and metal-air batteries are efficient solutions to reduce environmental issues and alleviate the energy demand. These technologies are primarily realized in the water cycle. The fuel cell has gained significant attention among the alternatives to fossil fuels due to its impressive conversion efficiency, reliable performance, fast startup, operation at low temperatures, and minimal emissions. The fuel cell converts chemical energy directly into electricity without the emission of pollutants. In other words, Hydrogen (H$_2$), which is generated by harvesting sustainable solar and wind energies, is used as input and results in power generation with absolutely zero emission.[4] So, fuel cells are the ideal choice for energy conversion and power sources for a healthy and sustainable future. The fuel cell relies on some of the electrochemical reactions such as hydrogen evolution reaction (HER,) ORR and oxygen evolution reaction (OER) reactions. However, the practical growth of fuel cells has been extensively hindered by the inherent thermodynamic limitation of the ORR.[5] To utilize the ideal performance of the fuel cell, it is necessary to hasten the kinetics of ORR. So, today's demand is to develop an active material that can serve as an efficient catalyst and would assist to accelerate the kinetics of the cathodic reaction. To date, Platinum (Pt) and Pt-based materials serve as a suitable catalyst for the reduction of $O_2$ at the cathode in the fuel cell.[6–8] Moreover, a sizable overpotential higher than 0.20 V is observed, even with the state-of-art Pt-based catalyst.[9] The utilization of Pt catalysts account for approximately 40-50% of the total cost associated with fuel cell stacks.[10] So, the cost of the fuel cell imposes serious limitations on the commercialization of the fuel cell. Moreover, the shortage of noble metals in the earth's crust and their insufficient stability towards the byproducts hinder their widespread applications in modern science and technology. Thus, mass production of such types of electrocatalysts is not feasible. Mainly the mass production of the ORR catalysts is impeded by the following three limitations: (1) the high cost:[10] currently, the Pt has been recognized as an effective catalyst in the fuel cells, which has a limited reserve and has resulted from the high cost of the fuel cells; (2) low performance:[11] another factor is the deficiency of the active sites on the catalyst. To actively use the efficiency of the fuel cells, it is necessary to fabricate a catalyst with a large number of active sites and specific selectivity; (3) poor stability:[12] another factor, which is the cause of the degradation of fuel cell performance, is the stability of the catalysts towards the byproducts formed during the operation of the fuel cell. The quest for enhanced catalysts for the ORR has sparked advancements in the identification of non-precious metal-based catalysts that exhibit superior catalytic activity compared to Pt. In this pursuit, researchers have explored various metals both in their pure state and through the process of alloying with other metals. Pt, Pd, Au, Ag, and Ni clusters are to be effective to catalyze oxygen reduction in their pure form.[13] Wang et al. have extensively documented the efficacy of synergistic Mn-Co catalysts in facilitating high-rate ORR.[14] Numerous recent studies have provided substantial evidence to support the exceptional catalytic properties of metal oxides and sulfides in the context of ORR, which gives efficiency as good as the state of art Pt/C catalyst.[15,16] Although metal catalysts have demonstrated excellent catalytic activity but considering the low availability of metal in the earth's crust and huge cost, the expansion of a cheap and efficient catalyst is a paramount need for realizing the practical applications of fuel cells.

As a typical earth-abundant material, transition-metal compounds such as transition metal oxide, metal chalcogenides, metal phosphides, metal carbides, metal nitrides, etc. have been widely used as promising catalysts for electrochemical mechanisms in fuel cells and energy conversion applications.[17] Among all the transition metal compounds, two-dimensional transition metal dichalcogenides (2D TMDs) have gained significant interest as a catalyst for electrochemical reactions. The unique properties of 2D TMDs which make them suitable for the catalyst are: (1) large surface area: the 2D TMDs have a high surface-to-volume ratio which provides a large density of active sites; (2) stability and thermal conductivity: as the 2D TMDs have only planar covalent bonding, they show excellent mechanical strength and remarkable resilience, and thermal conductivity enables efficient dispersion of heat generated during exothermic reactions; (3) tunable electronic properties: the electronic properties of the 2D TMDs materials can be tailored by creating defects, doping, stress, etc.[18] The electronic properties in turn tailor the catalytic performance of the materials. Owing to the above properties the 2D TMDs are good candidates for the electrocatalyst of electrochemical reactions. In the wide family of 2D TMDs, 2D $MoS_2$ has emerged as a highly investigated material due to its remarkable catalytic capabilities.[19] $MoS_2$ has been considered a promising non-noble HER catalyst but only the edge site shows high HER activity.[20,21] In other words, experimental and computational studies confirm that the electrocatalyst activity of $MX_2$ (where M is transition metal atoms and X is chalcogens) is closely correlated with the edge side while the basal plane with a large area is generally inactive.[22,23] To the best of our knowledge, the optimization of both active site and electronic properties of 2D $MX_2$ electrocatalysis is highly required. As we know the basal plane of 2D TMDs constitutes a large surface area, therefore if the inert basal planes could be activated, high electrocatalytic activity would be expected. To activate the inert basal plane and tailor the electronic properties, various strategies have been developed such as creating defects,[24,25] metal doping,[26,27] nonmetal substitution, strain engineering,[28] phase engineering, etc.[29] Among all the strategies, substitutional doping is considered a suitable technique as it provides long-term stability without degrading performance. The dopant atoms change the spin-density distribution around them, which would change the strength of $O_2$ adsorption and other reaction species. So, the ORR process can be performed on the surface of 2D TMDs near the appropriate doping atoms. For instance, Xiao et al. theoretically demonstrated the favorable catalytic activity of Co/Ni-doped 2D $MoS_2$ towards ORR.[30] Pumera et al. put forward the idea that introducing Fe and Mn dopants into the 2D $MoS_2$ can promote its catalytic activity in the context ORR.[31] Recently, Singh and Pakhira computationally analyzed that 2D monolayer Pt-doped $ZrS_2$ exhibits a superior catalytic activity towards ORR.[27] Moreover, the doping of $MoS_2$ by heteroatoms such as P and N exhibits moderate adsorption strength of oxygen molecules and enhances the ORR.[32,33] In a recent study by Tain and Tang , it was discovered that the addition of Ni or Co to materials such as 1T-$TiS_2$, 2H-$TiS_2$, 1T-$ZrS_2$, 1T-$NbS_2$, and 2H-$TaS_2$ results in a promising electrocatalytic activity for the ORR. The overpotential required for ORR on these doped materials was found to be within the range of 0.32-0.55 V, which is comparable to that of the most advanced electrocatalysts currently available (i.e., Pt).[34] It has been shown that the 2D monolayer $ZrSe_2$ is a new material of TMDs and has been successfully prepared experimentally and it has 1T phase as a stable structure.[35–37] Moreover, earlier predictions show that at low vacancy density, the $ZrSe_2$ and $ZrTe_2$ TMDs have a low value of hydrogen adsorption energy ($\Delta E^H$) and the value of $\Delta G$ is very close to that of Pt.[38]

In a recent theoretical study, Som and Jha showed an important observation regarding the catalytic activity of both the 2D single layer $ZrS_2$ and $ZrSe_2$ materials. They found that the edge sites of these materials, rather than the basal planes, play a crucial role in accelerating the HER. To further enhance the HER kinetics, the researchers decided to introduce the Nb, Pt, and W dopants in both the 2D $ZrS_2$ and $ZrSe_2$ materials. Interestingly, they discovered that the 2D Nb-doped systems, namely Nb-$ZrS_2$ and Nb-$ZrSe_2$, exhibited a robust metallic nature, which significantly boosted their catalytic activity towards HER. Furthermore, they observed that, on the basis of $\Delta G$ calculations, 2D monolayer Nb-$ZrSe_2$ has the best catalytic activity towards HER.[39] In addition, 2D $ZrS_2$ possesses important characteristics that make it a favorable choice. It is non-toxic, affordable, and exhibits strong resistance to corrosion. Furthermore, it has been successfully synthesized on a large scale and experimented with in the form of thin atomic layers.[40] In the context of sulfur/selenium-based $ZrS_2$ or $ZrSe_2$ TMDs, the presence of S or Se with a $2^-$ valence ($S^{2-}$ or $Se^{2-}$) enhances the reduction capacity, thereby promoting the activity of the ORR. Motivated by the extensive research attention and extraordinary catalytic activity of the Nb-doped $ZrSe_2$ toward HER, it is the fundamental interest to explore the more catalytic applications of 2D Nb-$ZrSe_2$ towards ORR. In this work, a DFT-D calculation is performed to explore the catalytic activity of the 2D monolayer Nb-$ZrSe_2$ towards ORR. In principle, we aimed to answer these questions: how the doping of Nb alters the structural and electronic properties of the 2D monolayer $ZrSe_2$? Do the 2D Nb-$ZrSe_2$ catalyst acts as an efficient ORR catalyst? If yes, which path it will follow to reduce the $O_2$? To answer these questions, we have analyzed the structural and electronic properties of 2D Nb-$ZrSe_2$ and examined the changes of Gibbs free energy ($\Delta G$) for all ORR intermediates at their minimum-energy condition. This study addresses some unique properties of the 2D monolayer Nb-ZrSe$_2$ and suggests that it can act as an efficient and useful catalyst for the ORR mechanism. Thus, this 2D Nb-ZrSe$_2$ can serve as an efficient cathodic catalyst in fuel cells and metal-air batteries.

**THEORY, METHODOLOGY, AND COMPUTATIONAL DETAILS**

The ORR has an essential role in the functioning of fuel cells. These electrochemical devices are designed to convert the chemical energy of the fuel, typically hydrogen (H$_2$), directly into electrical energy. The working principle of the fuel cell can be defined as follows:[41,42] hydrogen is produced by the electrocatalytic cell as fuel is brought into the anode side of the fuel cell. At the anode side, H$_2$ undergoes a reaction resulting in the generation of electrons and protons:

$$H_2 \longrightarrow 2H^+ + 2e^-$$

The protons (H$^+$) pass through the membrane, while the accompanying electrons travel through the electric circuit. They combine together at the cathode side of the fuel cell with oxygen. At the cathode, oxygen is reduced to water by reacting with H$^+$ and e$^-$ through the electrochemical reaction:[43,44]

$$2H^+ + \frac{1}{2}O_2 + 2e^- \longrightarrow H_2O$$

All these processes rely on some electrochemical reactions. Among them ORR is one of the reactions at the cathode which reduces oxygen molecule to water.

ORR reaction involves oxygen diffusion, i.e., the O$_2$ adsorption on the surface of the catalysts followed by the H$^+$ + e$^-$ transfer processes to reduce the O$_2$ into water. The oxygen reduction in acidic media mainly occurs through two different pathways:-[45,46] either a four-electron reduction pathway from O$_2$ to H$_2$O or two electron pathway from O$_2$ to H$_2$O$_2$.

Direct 4e$^-$ reduction:

$$O_2 + 4H^+ + 4e^- \longrightarrow 2H_2O \qquad E^o = 1.23 \text{ eV}$$

Indirect reduction:

$$O_2 + 2H^+ + 2e^- \longrightarrow H_2O_2 \qquad E^o = 0.68 \text{ eV}$$

$$H_2O_2 + 2H^+ + 2e^- \longrightarrow 2H_2O \qquad E^o = 1.77 \text{ eV}$$

The indirect 2e$^-$ pathway produces corrosive peroxide (H$_2$O$_2$) intermediates, which may corrode the catalytic surface, hence there would be concern about the stability of the fuel cell. Moreover, the formation of $H_2O_2$ hinders the process and kinetics of ORR reaction but also deteriorates polymer membrane by generating reactive radicals. Thus, four electron pathway is preferred and considered more efficient than that of the two-electron pathway. Based on how the O=O bond breaks, the 4e⁻ pathway can be divided into two different reaction pathways:[47] (1) $O_2$ decomposition pathway: in this path the $O_2$ is adsorbed on the surface of catalyst and followed by the cleavage of $O_2$ bond. Then, it involves a series of electron-proton transfer reactions to reduce $O_2$ into water. The detail reaction steps can be described as follows:

$$O_2 + * \longrightarrow O_2^*$$

$$O_2^* \longrightarrow 2O^*$$

(2) OOH dissociation path: the $O_2$* reacts with H⁺ and e⁻ coming from anode side to form OOH*. Then, cleavage of the O=O bond occurs into O* and OH*. After that, the hydrogenation of the O* occurs to form the next OH*. Finally, the removal of $H_2O$ occurs from the surface of the catalyst. The details can be described as follows:

$$O_2^* + H^+ + e^- \longrightarrow OOH^*$$

$$OOH^* \longrightarrow O^* + OH^*$$

$$O^* + H^+ + e^- \longrightarrow OH^*$$

$$OH^* + H^+ + e^- \longrightarrow H_2O$$

Where the asterisk * denotes the adsorption site. In summary, we can say that the ORR reaction basically relies on the cleavage of O=O bond and protonation of the ORR intermediates.

The catalytic potential of the 2D Nb-ZrSe₂ towards ORR is measured combinedly by computing and analyzing the adsorption energy with the values of ΔG of each intermediate species during the subject reaction. We used computational hydrogen electrode (CHE) method to calculate the adsorption energy and the values of ΔG of each intermediate. In this method, the energies are calculated under the standard conditions (pH=0, p=1 bar, T=298.15 K, and U=0 V). We have implemented the approach introduced by Nørskov et al. in our study. Their research showcased a significant finding that the chemical potential of H⁺ + e⁻ can be correlated with 1/2 $H_2$ in the gaseous state, which was determined using the standard hydrogen electrode.[9] Consequently, under the standard conditions, we can calculate the energy change of the reaction A + H → A + H$^+$ + e$^-$ by employing the reaction A + H → A + 1/2 H$_2$. The adsorption (or binding) energies (ΔE) presented in this study has been calculated as the energy difference between the energy of the model with adsorbed species [E$_{slab+adsorbate}$] and the energy of the catalytic model [E$_{slab}$] and the energy of the adsorbate [E$_{adsorbate}$] according to the following equation:[48]

$$\Delta E = E_{slab+adsorbate} - E_{slab} - E_{adsorbate}$$

Negative adsorption energy indicates that the adsorbed oxygen intermediates are attached to the catalytic surface stably. In other words, a negative adsorption energy signifies that the adsorbate is likely to bind energetically to the catalyst's surface, indicating a favorable interaction between them. Therefore, negative adsorption energy is favorable for elementary reactions over the catalytic surface. The value of ΔG in each reaction step was evaluated as:[49,50]

$$\Delta G = \Delta E + \Delta E_{ZPE} - T\Delta S$$

Where ΔE denotes the adsorption energy obtained from the DFT-D calculations of the equilibrium structures and ΔE$_{ZPE}$ and ΔS represent the change in zero-point energy and entropy correction, respectively. T denotes the room temperature 298.15 K in this calculation.

For all the systems studied here (i.e., ZrSe$_2$, Nb-ZrSe$_2$, and ORR intermediates), the equilibrium atomic configuration, structures, electronic properties, and energetics were obtained by using the spin-unrestricted B3LYP DFT framework implemented in CRYSTAL17 suite code.[51] To account the weak non-bonding van der Waals (vdW) interactions between the reactants or intermediates and the interface, we incorporated the damped vdW dispersion correction (-D3) within the framework of density functional theory (DFT), specifically known as DFT-D3. This approach was developed by Grimme and co-workers.[52] A 2x2 supercell of the 2D monolayer structure of the pristine ZrSe$_2$ was considered and one Zr atom was substituted by one Nb atom to form the 2D monolayer Nb-ZrSe$_2$ structure, and all the ORR reaction steps on the surface of the 2D Nb-ZrSe$_2$ were investigated in the present study. We have utilized the Gaussian types of atomic basis sets to define the atomic orbitals, which have been found to yield more precise outcomes compared to plane wave basis sets.[53,54] In simpler terms, although the computational procedure differs from the plane wave code (i.e., VASP, Quantum Espresso), both approaches are yielding identical results. In the realm of hybrid density functionals, localized Gaussian basis set codes are inherently well-suited for addressing the Hartree-Fock (HF) component of the outcome.[51,55] For the Zr, Se, O, H, and Nb atoms, triple-zeta valance polarized (TZVP) Gaussian basis set were used in the present calculations.[56,57] We employed full optimization process i.e., all atomic coordinates and lattice parameters were allowed to relax during geometry optimization. The convergence criterion for the self-consistent field (SCF) was established as a total energy difference of $10^{-7}$ a.u. between two consecutive iterations. The threshold controlling of the coulomb exchange integral calculation were controlled by the five-threshold set 7, 7, 7, 7 (ITOL1 to ITOL4) and 14(ITOL5) for both the geometry optimization and electronic property calculations. The CRYSTAL17 code utilizes the default threshold for geometry optimization on all atoms, employing specific values for maximum and RMS force (0.000450 a.u. and 0.00300 a.u., respectively) and maximum and RMS displacement (0.001800 a.u. and 0.001200 a.u., respectively). To ensure accurate 2D monolayer i.e., SLAB calculations using the CRYSTAL17 code, we have established a separation of 500 Å in the normal surface direction of the 2D monolayer Nb-ZrSe$_2$ along Z-direction where is no symmetry.[27,58] This distance effectively prevents any spurious interactions between the periodic images of the monolayer slab. Our current calculations indicate that this separation is sufficient to mitigate any undesired influences. To perform the integration within the Brillouin zone, we have used 16x16x1 k-mesh in the Monkhorst pack scheme[59] for the pristine 2D ZrSe$_2$, Nb -ZrSe$_2$, and all the reaction intermediates for both the geometry optimization and electronic property calculations. The electronic band structure was constructed along high symmetry k-path direction ***Γ-M-K-Γ*** of the corresponding irreducible Brillouin zone. All the equilibrium structures with the images were plotted by using VESTA software.[60]

**RESULTS AND DISCUSSION**

The equilibrium structure of the 2D monolayer ZrSe$_2$, which is the main object of this present work, belongs to 1T phase (i.e., one atom of Zr is octahedrally coordinated by six atoms of Se) as shown in Figure 1-2. By using VESTA software, a 2D monolayer ZrSe$_2$ primitive cell was computationally designed, and a model system has been developed for further studies, and the structure was optimized by using the first principle-based periodic dispersion-correction hybrid

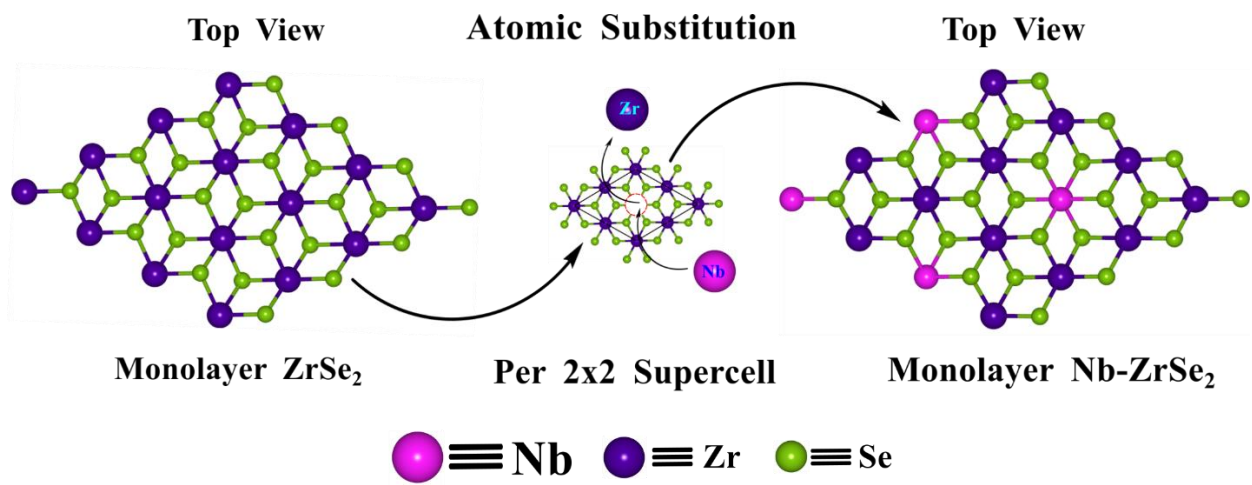

**Figure 1.** Representation of the 2D monolayer ZrSe$_2$ and 2D monolayer Nb-ZrSe$_2$.

DFT (DFT-D3, noted by B3LYP-D3) method as shown in Figure 1. The structure is a trigonal 2D slab system with $P\bar{3}m1$ layer group symmetry, and the computed unit cell parameters i.e., the lattice constants are about a = b = 3.75 Å of the 2D pristine monolayer ZrSe$_2$ which is well consistent with the previously reported values.[39,40] Each unit cell consists of one Zr and one Se atom as shown in Figure 1-2. In Figure 2, the computed equilibrium average bond length Zr-Se between the Zr and Se atoms is about 2.68 Å. The corresponding bond angles ∠ZrSeZr and ∠SeZrSe are estimated to be 88.99° and 91.00°, respectively. Then, the doped system is constructed by substituting one Zr atom with one transition metal Nb per 2x2 supercell of the 2D monolayer pristine ZrSe$_2$ as illustrated in Figure 1. The doping was defined by the following equations:

$$\theta = \frac{number\ of\ doping\ atom\ per\ 2x2\ supercell}{number\ of\ total\ metal\ atom\ in\ 2x2\ supercell}$$

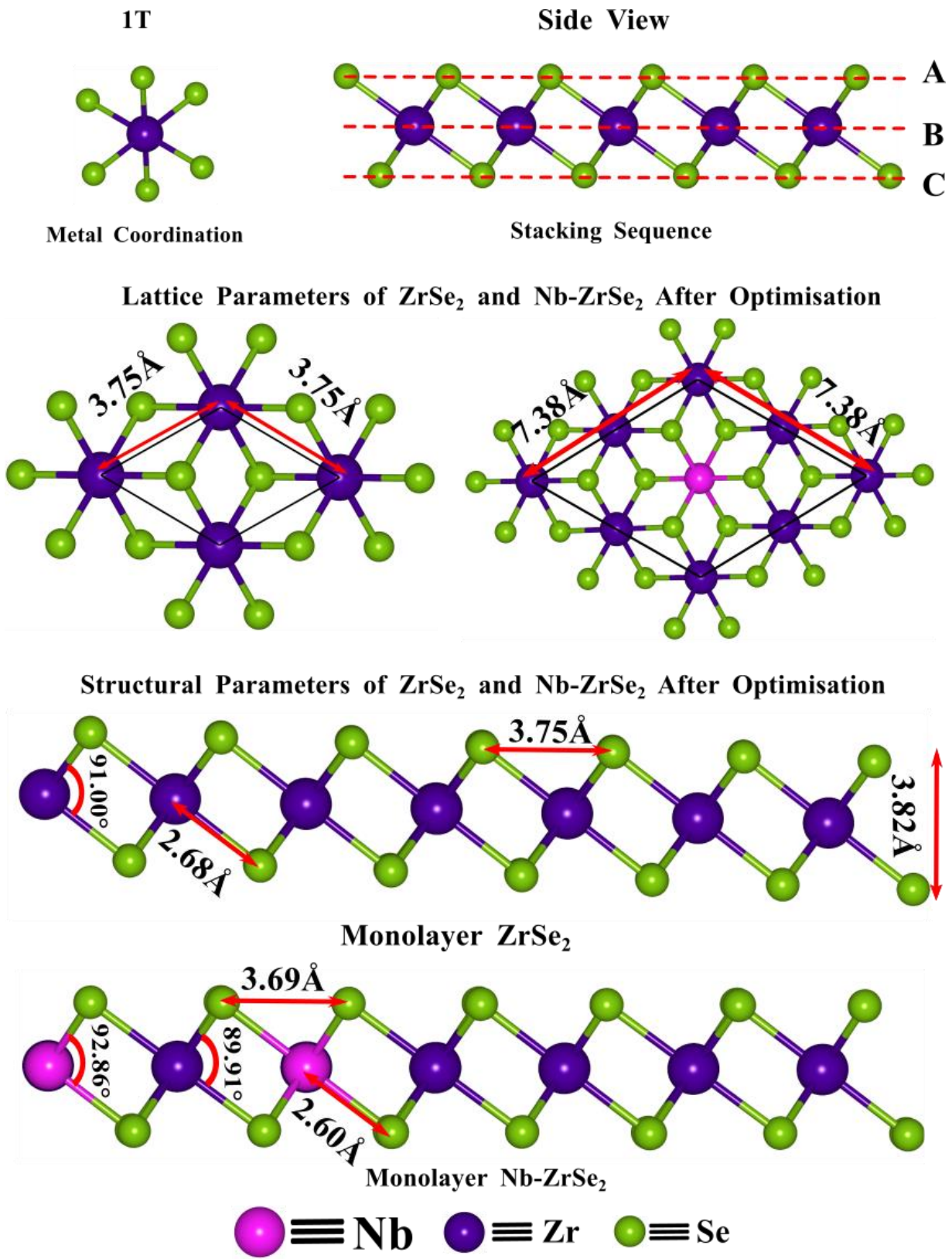

**Figure 2.** Metal coordination, stacking sequence, unit cell, and structural properties of ZrSe$_2$ and Nb-ZrSe$_2$.

So, the Nb dopant atom amounts to 25% (only metal atom doping) atomic substitution in per 2x2 supercell of the 2D ZrSe$_2$. The 2D monolayer structure of Nb-ZrSe$_2$ consists of one Nb, three Zr, and eight Se atoms per one unit cell as represented in Figure 2. The 2D Nb-ZrSe$_2$ system is then fully relaxed with respect to both atomic coordinates and cell parameters by the same level of DFT-D method during optimization. The equilibrium structure of the 2D Nb-ZrSe$_2$ material belongs to the *P1* (No. 1) layer group symmetry. The equilibrium lattice parameters of the 2D Nb-ZrSe$_2$ were found to be a = b = 7.38 Å as represented in Figure 2. The equilibrium average Zr-Se and Nb-Se bond lengths were found to be 2.67 Å and 2.60 Å, respectively. The equilibrium bond angles ∠ZrSeZr, ∠NbSeZr, ∠SeNbSe, and ∠SeZrSe were found to be 87.51°, 88.61°, 92.86°, and 92.86°, respectively, as represented in Figure 2. All the fully relaxed structural parameters of 2D monolayer ZrSe$_2$ and Nb-ZrSe$_2$ material have been reported in Table 1. The ORR mechanism involves the movements of the electrons i.e., the transfer or transportation of electrons is a key process during the electrochemical ORR mechanism. Previous studies have shown that the electronic properties (i.e., electronic band gap, band structure and total density of states) of catalysts have a significant impact on their catalytic efficiency. The electronic properties have been analyzed by computing and observing the electronic band structure and total density of states (DOSs) of both the 2D monolayer $ZrSe_2$ and Nb-$ZrSe_2$ materials. Analysis of the electronic properties can be useful to obtain information about electron distribution on the catalytic surface which is useful to understand fully the catalytic performance of the catalyst. The electronic properties of the 2D monolayer $ZrSe_2$ and Nb-$ZrSe_2$ materials have been obtained by employing the same DFT-D theory. The electronic band structure of the pristine $ZrSe_2$ were plotted in the highly symmetric k-path direction **Γ-M-K-Γ** with respect to vacuum. The highly symmetric path has been shown in Figure 3a, and the equilibrium electronic structure of the 2D monolayer $ZrSe_2$ material is shown in Figure 3b (top and side view). The Fermi energy level ($E_F$) of the pristine $ZrSe_2$ was found at the energy value of -5.34 eV as shown by the dotted line in Figure 3c-d. The 2D single layer $ZrSe_2$ material shows semiconducting characteristics with the indirect band gap of about 1.48 eV as represented in Figure 3c. The DOSs has been computed by the same level of DFT-D theory, and the profile of DOSs supports the band gap of the 2D pristine $ZrSe_2$ which was obtained by the electronic band structures calculations as shown in Figure 3d. The calculated electronic band gap of the pristine 2D $ZrSe_2$ is consistent with the previously reported values.[39,40] The electronic property calculation depicts that the 2D pristine monolayer $ZrSe_2$ has a noticeable indirect band gap of 1.48 eV, thus the utilization of this substance as a catalyst to improve the rate of the ORR is not feasible. Doping Nb in the pristine 2D monolayer $ZrSe_2$ (Nb-$ZrSe_2$) disrupts its symmetry and introduces additional unsaturated electrons. The equilibrium structure with the electronic properties of the 2D Nb-$ZrSe_2$ have been calculated at its equilibrium geometry by using the same level of the DFT-D theory as depicted in Figure 4a-d. This alteration leads to an augmented charge density in the basal plane of the 2D monolayer Nb-$ZrSe_2$ resulting in enhanced catalytic activity towards ORR. We have drawn the electronic band structure of the 2D Nb-$ZrSe_2$ along the highly symmetric k-path direction **Γ-M-K-Γ** like the pristine 2D monolayer $ZrSe_2$ material with respect to the vacuum for comparison. The $E_F$ of 2D Nb-$ZrSe_2$ was found at -4.87 eV as represented by the dotted line in Figure 4b-c. The results from our present DFT-D analysis demonstrate that the introduction of Nb doping in the pristine 2D monolayer $ZrSe_2$ causes the energy gap to disappear, stemming in a zero-band gap where the bands are overlapped around the $E_F$ making it conductor. In other words, the 2D Nb-ZrSe$_2$ has metallic characteristics as shown in Figure 4b. The zero-band gap indicates easier electron transfer from valence band to the conduction band. To support this, we computed the DOSs calculation using the same level of DFT-D theory. The DOSs profile show that there is sufficient carrier concentration about the $E_F$ as shown in Figure 4c. An increased concentration of electrons near the $E_F$ enhances electronic conductivity, thereby facilitating faster electron transfer to the reactants during the process of ORR. The carrier density around the $E_F$ mostly comes from the dopant (Nb) **d**-subshell electron density of states. Based on our analysis of partial density of states (PDOS) calculations, we have successfully established the influence of the **d**-subshell electron density of states of the Nb atom on the carrier concentration around the $E_F$, as depicted in Figure 4d. The contributing component of the **d**-subshell electron density of the Nb atom in the total DOSs has been computed to examine the conducting properties of the material where the **d**-subshells electron density of states of the Nb atom follow the locus of the total DOSs of the 2D monolayer Nb-ZrSe$_2$ material. In other words, this indicates that the **d**-subshell electrons of the Nb atom regulates and tunes the electronic properties of the 2D monolayer Nb-ZrSe$_2$ material which may enhance the electrocatalytic activities of the material.

**Table 1.** The equilibrium structural properties of 2D monolayer ZrSe$_2$ and 2D monolayer Nb-ZrSe$_2$.

| System | Lattice parameters (Å) | Interfacial angle in degree | Layer group and symmetry | Average bond length | | References |
| --- | --- | --- | --- | --- | --- | --- |
| | | | | Zr-Se (Å) | Nb-Ze (Å) | |
| ZrSe$_2$ monolayer | a=b=3.75 | a=β=90°<br>γ=120° | $p\bar{3}m1$ | 2.68 | - | This work |
| ZrSe$_2$ monolayer (previously reported) | a=b=3.77 | a=β=90°<br>γ=120° | $p\bar{3}m1$ | 2.70 | - | 39,40 |
| Nb-ZrSe$_2$ monolayer | a=b=7.38 | a=β=90°<br>γ=120° | $P1$ | 2.67 | 2.60 | This work |

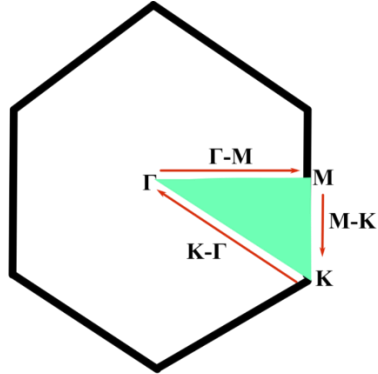

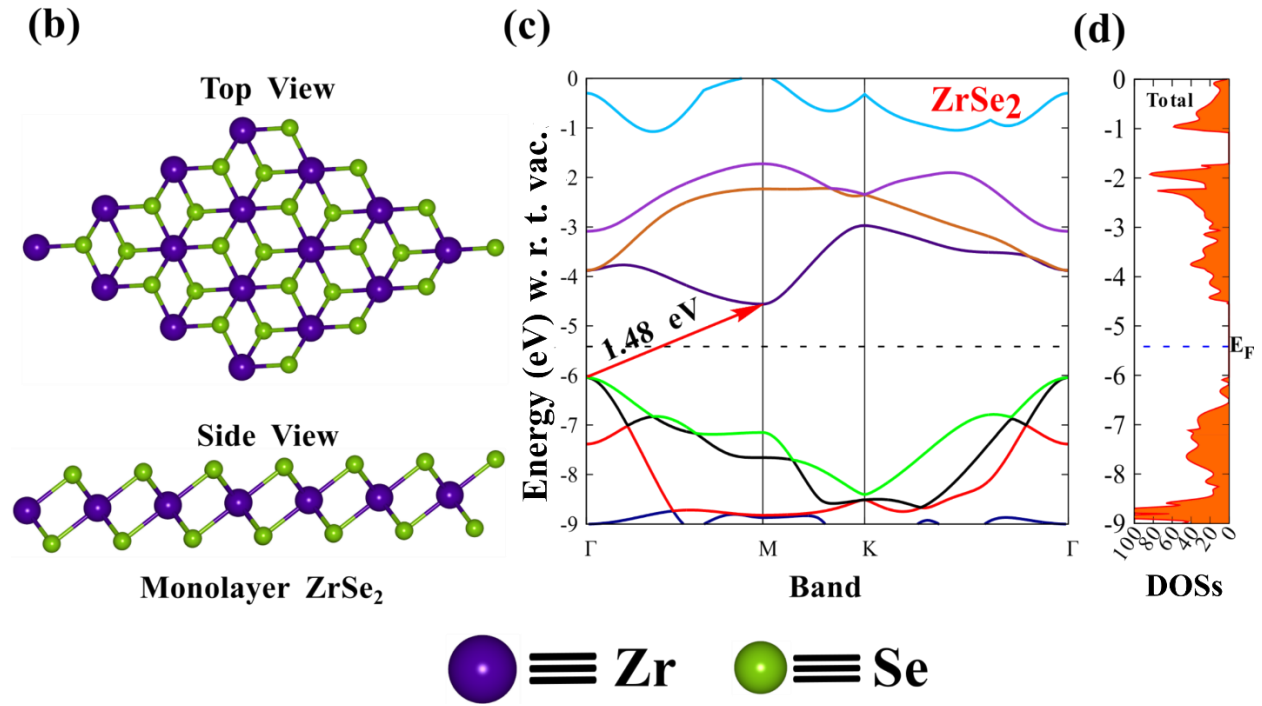

**Figure 3.** (a) Brillouin zone with high symmetry k-points, (b) top and side view of equilibrium structure of monolayer ZrSe$_2$, (c) band structure, and (d) total DOSs of monolayer ZrSe$_2$.

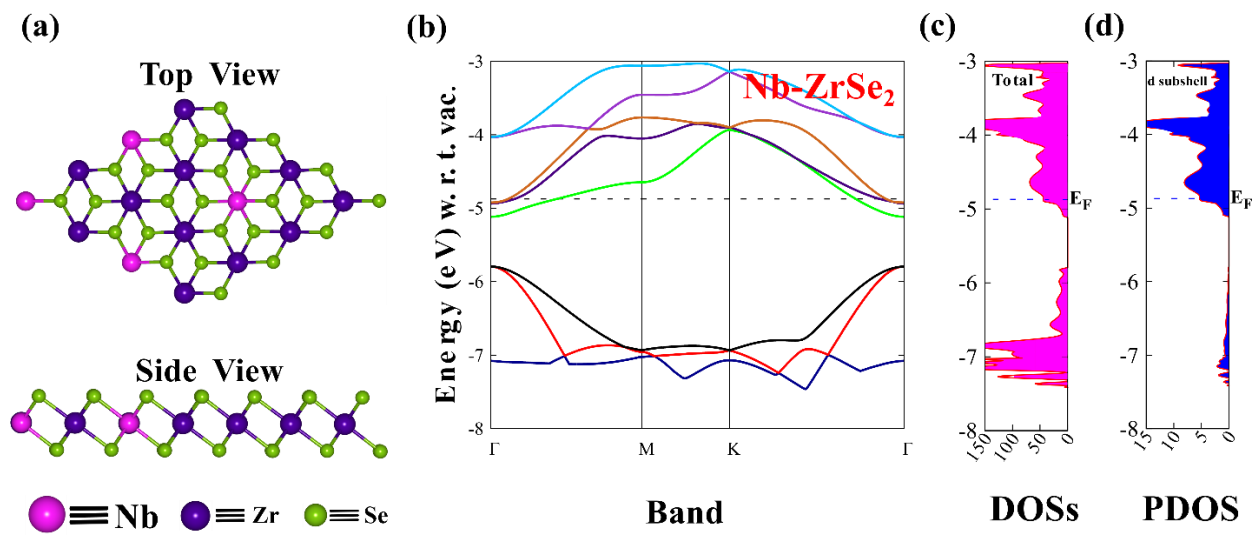

**Figure 4.** (a) Top and side view of equilibrium structure of monolayer Nb-ZrSe$_2$, (b) band structure of monolayer Nb-ZrSe$_2$, (c) total DOSs of Nb-ZrSe$_2$, and (d) contribution of d subshell of Nb in the total DOSs.

The distinctive electronic characteristics of the 2D monolayer Nb-ZrSe$_2$ can be summarized as follows: (1) the rapid transfer of charges is facilitated by the presence of continuous band states in close proximity to the $E_F$; (2) enhanced electron mobility can be observed due to the significant density of states surrounding the $E_F$; (3) the metallic properties of the 2D monolayer Nb-ZrSe$_2$ contribute to accelerate charge transfer leading to improve the ORR activity.

Considering the previous studies of ORR mechanism on TMDs, the surface chalcogen atomic layer takes part in the reduction process of the molecular oxygen ($O_2$). Recently Singh and Pakhira investigated the ORR mechanism on the surface chalcogen (S) atomic layer of the 2D monolayer Pt-ZrS$_2$ using DFT-D methods.[27] So, in the case of TMDs, it can be concluded that the surface chalcogen atomic layer is the active center of the catalyst during electrochemical ORR mechanism. In this work, we have investigated the ORR mechanism in an acidic media, where the protons ($H^+$) and electrons ($e^-$) transfer take place simultaneously on the ORR adsorbates to reduce the $O_2$. We considered the Se site as an active site (near the Nb-doped region) for the ORR adsorbates including $O_2^*$, $2O^*$, $OOH^*$, $O^*$, and $OH^*$.

**Table 2.** Adsorption energy ($\Delta E$ in eV) of various reaction intermediates involved in both the dissociative and associative pathways of ORR on the surface of the 2D monolayer Nb-ZrSe$_2$ material.

| Various reaction steps involved in dissociative/associative mechanism | $\Delta E$ (eV) |
|---|---|
| [Nb-ZrSe$_2$+O$_2$ → O$_2^*$_Nb-ZrSe$_2$] | 0.15 |
| [O$_2^*$_Nb-ZrSe$_2$ → 2O$^*$_Nb-ZrSe$_2$] | 1.94 |
| [O$_2^*$_Nb-ZrSe$_2$+H$^+$+e$^-$ → OOH$^*$_Nb-ZrSe$_2$] | -0.18 |
| [OOH$^*$_Nb-ZrSe$_2$+H$^+$+e$^-$ → O$^*$_Nb-ZrSe$_2$+H$_2$O] | -1.47 |
| [2O$^*$_Nb-ZrSe$_2$+H$^+$+e$^-$ → O$^*$_OH$^*$_Nb-ZrSe$_2$] | -2.65 |
| [O$^*$_OH$^*$_Nb-ZrSe$_2$+H$^+$+e$^-$ → O$^*$_Nb-ZrSe$_2$+H$_2$O] | -0.95 |
| [O$^*$_Nb-ZrSe$_2$+H$^+$+e$^-$ → OH$^*$_Nb-ZrSe$_2$] | -1.59 |
| [OH$^*$_Nb-ZrSe$_2$+H$^+$+e$^-$ → Nb-ZrSe$_2$+H$_2$O] | -1.54 |

The ORR mechanism was started by analyzing the interaction of $O_2$ with the active site Se of the 2D monolayer Nb-ZrSe$_2$. The adsorption of $O_2$ with the active site is the crucial step to initiate the ORR process and significantly affects the catalytic activity of the material. Here, we calculated the adsorption energy of the $O_2$ molecule by taking the difference between the electronic energy of the $O_2$ absorbed Nb-ZrSe$_2$ denoted by $O_2$*_Nb-ZrSe$_2$ with electronic energy of isolated 2D Nb-ZrSe$_2$ and gaseous state of the $O_2$ molecule. The adsorption energy of $O_2$ should not be too positive or too negative for a good electrocatalyst.[27,58,61,62] The higher positive adsorption energy value between $O_2$ and the active site (called $O_2$-active site interaction) depicts that it is not able to capture the $O_2$ molecule. In other words, the higher and positive value of $O_2$ adsorption energy (around the active center) leads to poor absorption of $O_2$ molecule, so it is very difficult to start the ORR process. A very negative adsorption energy value between $O_2$ and catalyst is not ideal for ORR mechanism because a very strong $O_2$-active site interaction may lead to occupation of active site by $O_2$ molecule and hinders the further reaction steps.[27,58,61,62] Thus, a very large positive and negative values of the $O_2$ adsorption energy around the active site of the catalyst diminish the electrocatalytic activity of the material towards the ORR process. So, the adsorption energy of $O_2$ should be optimum to efficiently start the ORR on the catalytic surface. Based on the above discussion, we computed the $O_2$ activation energy, which could be good descriptor of the catalytic activity of the 2D monolayer Nb-ZrSe$_2$ towards ORR. In other words, the determination of the optimal value of $O_2$ adsorption energy on the Nb-ZrSe$_2$ surface plays a crucial role in initiating the ORR process. For an ideal condition, the adsorption energy of $O_2$ on the catalytic surface should be close to zero, but which is not the case in real life. Noble metals, such as Pt are considered the best catalyst till date for electrochemical reactions especially for ORR. So, for practical applications the $O_2$-active center interaction energy for a good catalytic candidate should be close to or slightly weaker than that of the Pt. Chen et al. reported that the Pt-based catalyst shows good ORR reactivity, therefore, can be used as a reference to evolute the catalytic activity of other materials. The $O_2$ adsorption energies on the Pt(111) and Pt(100) surfaces were found to be -0.69 eV and -1.10 eV, respectively.[6,63] It was found in the present calculations that $O_2$ adsorption energy on the surface of 2D monolayer Nb-ZrSe$_2$ was about -0.15 eV which is smaller than that of the $O_2$-active site interaction energy of Pt. Thus, this negative and small adsorption energy of $O_2$ suggests that the 2D monolayer Nb-ZrSe$_2$ can be energetically favorable for electrochemical ORR to proceed further, which indicates that the 2D monolayer Nb-ZrSe$_2$ material would show an efficient catalytic activity towards ORR.

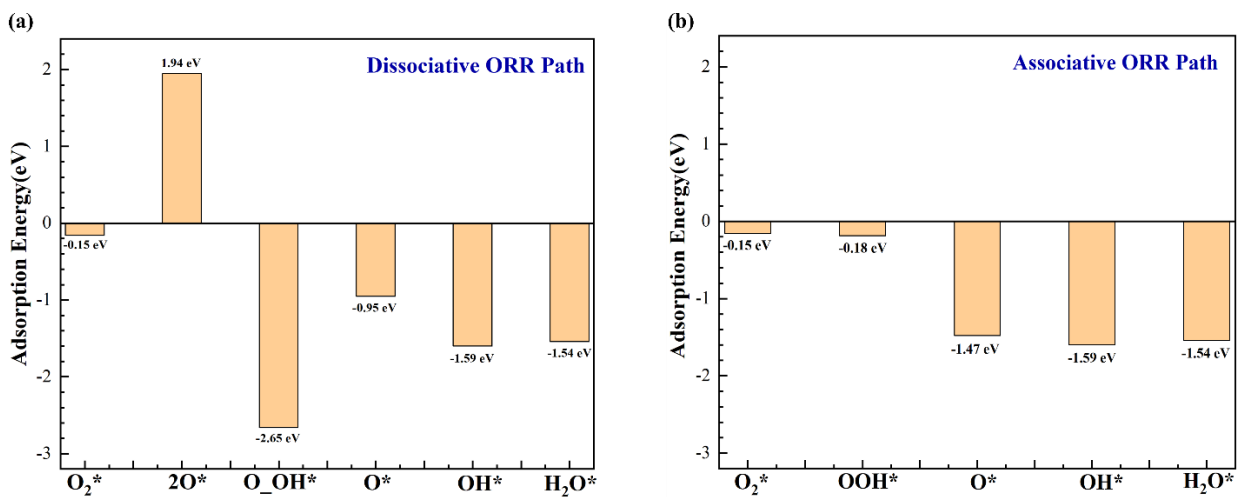

**Figure 5.** Graphical representation of adsorption energy of various ORR intermediates in (a) dissociative, and (b) associative mechanisms.

Now the removal of water ($H_2O$) molecule is the last step of ORR mechanism. After the removal of $H_2O$ molecule, the new ORR cycle of ORR mechanism starts. So, suitable $H_2O$ adsorption energy is essential during the ORR mechanism. The adsorption energy of $H_2O$ on catalytic site should be small. The adsorption energy of $H_2O$ molecule was found to be -1.54 eV on the Se-site as reported in Table 2. The relatively low adsorption energy observed on the Se site suggests that it is well-suited for the ORR mechanism. This means that the Se site remains accessible for the subsequent cycles of the ORR. Moreover, the favorable adsorption energies of $O_2$ and $H_2O$ on the Se site of 2D Nb-ZrSe$_2$ indicates that this material has the potential to serve as an effective catalyst for converting $O_2$ into $H_2O$ during the reduction process.

Now, after the adsorption of $O_2$ molecule, which is the initial step of ORR mechanism, ORR can proceed via two ways. The path of the ORR mechanism depends on how the O=O bond cleaves. The first way is called associative path, which involves both the proton and electron transfer simultaneously to form OOH species, then followed by the successive protonation steps to reduce $O_2$. The second path is known as dissociative path, which involves the O=O bond breaking after the adsorption of $O_2$ molecule into two O atoms on the active site of the catalytic surface. In this work, we have studied both the possible pathways of ORR mechanism on the surface of the 2D monolayer Nb-ZrSe$_2$. We computed the adsorption energy of the reaction intermediates with the active site Se of the 2D Nb-ZrSe$_2$ for both associative and dissociative reaction pathways. The adsorption energy for the hydrogenation of $O_2^*$ into OOH* species during associative reaction was found to be -0.18 eV obtained by the DFT-D calculation as represented in Figure 5. The adsorption energy of dissociative pathway (i.e., dissociation of the adsorbed $O_2^*$ into 2O*) was found to be 1.94 eV, i.e., the dissociation energy of $O_2^*$. According to computed adsorption energy, the associative pathway is more energetically favorable than that of the dissociative pathway. The adsorption energy for associative pathway ($O_2^* \rightarrow OOH^*$) indicates that the OOH* species is stably absorbed on the Se site of the 2D Nb-ZrSe$_2$. As the adsorption energy of associative path (-0.18 eV) is close to zero i.e., very small and negative, therefore we consider the associative ORR path would be thermodynamically favorable and feasible. Thus, we believe that associative reaction path is energetically more favorable and efficient for catalyzing $O_2$ molecules. The associative reaction path is represented as: $O_2^* + H^+ + e^- \rightarrow OOH^*$. The large positive adsorption energy of dissociation of $O_2$ molecule to the atomic O indicates that this step is not stable and would be greatly suppressed by the associative 4e$^-$ mechanism. A positive adsorption energy implies that the dissociation of $O_2$ is not energetically favorable, which hinders the progression of subsequent reaction steps. So, the whole ORR mechanism will proceed as $O_2^* \rightarrow OOH^* \rightarrow O^* + H_2O \rightarrow OH^* \rightarrow H_2O$. The adsorption energy of all ORR intermediates has been calculated by the same level of theory and reported in Table 2. The graphical representation of adsorption energy of all intermediate steps has been shown in Figure 5a-b. For all the intermediates states involved in the associative path, the adsorption energy is negative and optimum, revealing highly exothermic characteristics and thermodynamically favorable. We also computed the adsorption energy for all the ORR intermediates which occurs in the dissociative ORR pathway as represented in Figure 5a. Adsorption energy of all reaction intermediates involved in dissociative path has been reported in Table 2.

From the above discussion, we can conclude that the ORR mechanism on the surface of the 2D monolayer Nb-ZrSe$_2$ material will follow the 4e$^-$ associative mechanism. To further analyze the catalytic activity of the 2D monolayer Nb-ZrSe$_2$ towards ORR, we have computed the values of ΔG for each intermediate reaction steps. In this work, we have calculated the values of ΔG by computing the harmonic vibrational frequency of each intermediate step of ORR mechanism using the CRYSTAL17 code. We will discuss the ΔG, structural properties, and electronic properties of each reaction intermediate step involved in both associative and dissociative mechanism during the ORR.

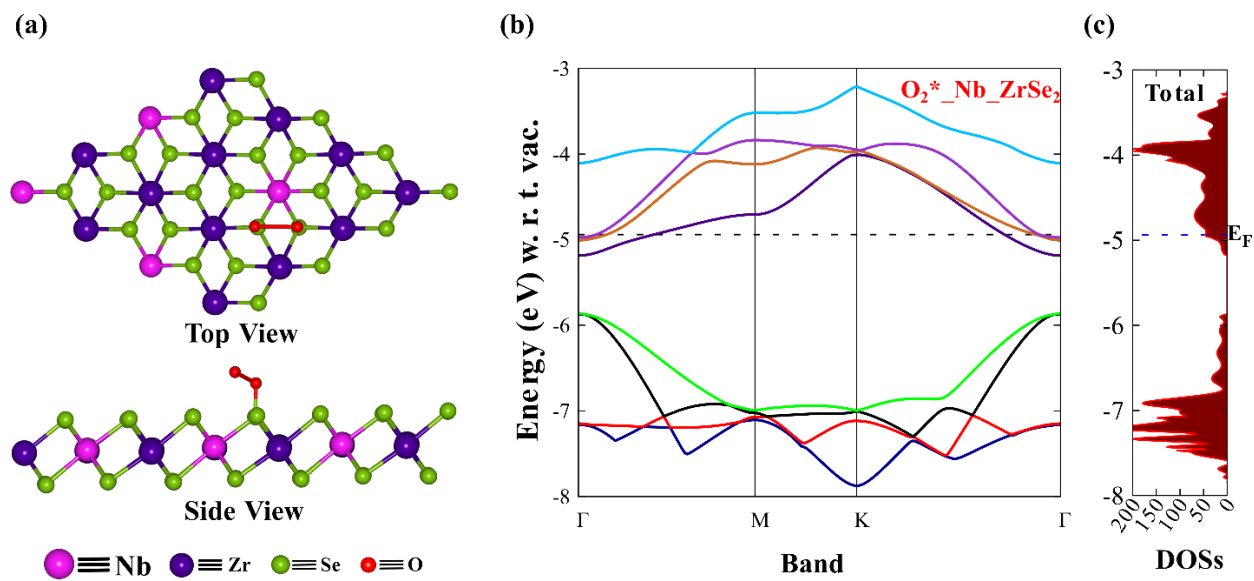

**Figure 6.** (a) Top and side view of equilibrium structure of $O_2^*\_Nb\text{-}ZrSe_2$, (b) band structure, and (c) total DOSs of $O_2^*\_Nb\text{-}ZrSe_2$.

**Reaction intermediate steps involved in associative mechanism:**

**I$^{st}$ step:** The first step of the ORR mechanism, irrespective of the path, is the adsorption of $O_2$ onto the reaction site of the 2D monolayer Nb-ZrSe$_2$ slab. To model this step, we placed an $O_2$ molecule on the surface near to the Nb atom in the 2D Nb-ZrSe$_2$ at a distance 1.60 Å which is equal to the equilibrium bond length of the Se-O. The same DFT-D method has been employed to perform the geometry optimization of the 2D Nb-ZrSe$_2$ with an $O_2$ molecule attached to the Se site (represented as $O_2^*\_Nb\text{-}ZrSe_2$). The equilibrium structure of $O_2^*\_Nb\text{-}ZrSe_2$ has been represented in Figure 6a (top and side view). We analyzed and calculated the intrinsic electronic characteristics of the equilibrium structure $O_2^*\_Nb\text{-}ZrSe_2$. In order to gain insights into the electronic properties, we calculated the electronic band structure and DOSs as depicted in Figure 6b and 6c, respectively. We plotted a total number of eight bands around the $E_F$, which are sufficient for collecting the electronic behavior of the $O_2^*\_Nb\text{-}ZrSe_2$. We calculated the band structure of $O_2^*\_Nb\text{-}ZrSe_2$ along the *Γ-M-K-Γ* high symmetry k-path direction with respect to vacuum, which is consistent to the 2D Nb-ZrSe$_2$ band structure with the band pathway along the same k-vector direction for comparison. The $E_F$ was found at -4.94 eV as represented by the dotted line in Figure 6b-c. From the band structure calculations, we observed that some of the electronic bands overlap around the E$_F$. Thus, the band structure of O$_2$*_Nb-ZrSe$_2$ predicts the conducting nature of the O$_2$*_Nb-ZrSe$_2$ system. To confirm the conducting nature of O$_2$*_Nb-ZrSe$_2$, we computed and analyzed the DOSs of O$_2$*_Nb-ZrSe$_2$. According to the DOSs profile, it appears to be an ample amount of electron density surrounding the E$_F$ as shown in Figure 6c. Therefore, O$_2$*_Nb-ZrSe$_2$ has metallic characteristics. After optimization, the equilibrium Se-O and O-O bond lengths were found to be 1.67 Å, and 2.20 Å, respectively. We observed an increased bond length of the O-O bond as compared to the free $O_2$ molecule bond length (1.21 Å) by the amount of 0.99 Å after the adsorption onto surface of the Nb-$ZrSe_2$ as represented in Figure 12. The increment in the bond length indicates that the $O_2$ molecule is activated and would be available for the next ORR steps. The optimized equilibrium structure of the 2D monolayer $O_2$*_Nb-$ZrSe_2$ belongs to the layer group symmetry *P1* (layer group symmetry number is no.1), with the optimized equilibrium lattice constants within computed by the DFT-D method are a = 7.37 Å, b = 7.38 Å, and angle between a and b i.e., γ = 120.01° as reported in Table 3. The value of ΔG during the reaction $O_2$ + * → $O_2$* was found to be -0.10 eV. The negative value of ΔG indicates that the reaction is exothermic. Thus, negative, and small value of ΔG suggests that the adsorption of $O_2$ on active site is thermodynamically stable and kinetically feasible for the ORR mechanism.

**Table 3.** Equilibrium lattice parameters and structural parameters of various ORR intermediates.

| Reaction steps | Lattice parameters (Å) | Interfacial angle (°) | Layer group and symmetry | Electronic band gap (eV) | Average bond length (Å) | | | |
|---|---|---|---|---|---|---|---|---|
| | | | | | Zr-Se | Nb-Se | Se-O | Se-OH |
| Nb-ZrSe$_2$ | a=7.38, b=7.38 | α=β=90° γ=120° | P1 | 0 | 2.67 | 2.60 | - | - |
| O$_2$*_Nb-ZrSe$_2$ | a=7.37, b=7.38 | α=β=90° γ=120.01° | P1 | 0 | 2.67 | 2.58 | 1.69 | - |
| 2O*_Nb-ZrSe$_2$ | a=7.39, b=7.35 | α=β=90° γ=120.36° | P1 | 0 | 2.66 | 2.60 | 1.67 | 1.67 |
| OOH*_Nb-ZrSe$_2$ | a=7.39, b=7.40 | α=β=90° γ=120.20° | P1 | 0.66 | 2.66 | 2.66 | 2.00 | - |
| O*_OH*_Nb-ZrSe$_2$ | a=7.39, b=7.43 | α=β=90° γ=120.32° | P1 | 1.39 | 2.68 | 2.61 | 1.72 | 1.77 |
| O*_Nb-ZrSe$_2$ | a=7.37, b=7.41 | α=β=90° γ=120.13° | P1 | 0 | 2.67 | 2.60 | 1.67 | - |
| OH*_Nb-ZrSe$_2$ | a=7.38, b=7.40 | α=β=90° γ=120.10° | P1 | 0.66 | 2.65 | 2.61 | 1.87 | - |

**II$^{nd}$ step:** The next step (in the 4e$^-$ transfer associative reaction mechanism of the subject reaction) is the hydrogenation of adsorbed activated oxygen molecule around the Se site near the Nb atom. To model this ORR intermediate, we have placed one H atom near to the one O atom and the present DFT-D study has found that the equilibrium O-H distance is about 0.97 Å from the outer O atom in the $O_2$*_Nb-$ZrSe_2$. Here, we assumed that a H$^+$ coming from the anode side through proton exchange membrane and an e$^-$ coming through external circuit reacts with the adsorbed $O_2$ molecule to form OOH (noted by OOH*_Nb-$ZrSe_2$). The full geometry optimization (i.e., atomic coordinates and lattice parameters) of the so formed system OOH*_Nb-$ZrSe_2$ reaction intermediate was performed by employing the same DFT-D method, and the equilibrium structure of the OOH*_Nb-$ZrSe_2$ intermediate is shown in Figure 7a (top and side view). We computed and examined the electronic properties of the OOH*_Nb-$ZrSe_2$ by using the same DFT-D method. The band structure was plotted along the high symmetric k-path ***Γ-M-K-Γ*** as shown in Figure 7b. We plotted eight bands around the $E_F$, which are sufficient for collecting the electronic behavior of the OOH*_Nb-$ZrSe_2$ followed by our previous calculation. The $E_F$ was found at -5.42 eV shown by the dotted line in Figure 7b-c. The band structures of the OOH*_Nb-$ZrSe_2$ depicts the semiconducting nature of the OOH*_Nb-$ZrSe_2$ reaction intermediate with an indirect band gap of 0.66 eV. The conduction band minima and the valance band maxima of the OOH*_Nb-$ZrSe_2$ are located as Γ- and M- points, respectively. The DOSs corresponding to the band structures of the OOH*_Nb-ZrSe$_2$ were also calculated and plotted by using the same DFT-D method as shown in Figure 7c. The DOSs plot is well consistent with the band structures and this computation supports the band gap of the OOH*_Nb-ZrSe$_2$ intermediate. Our DFT-D calculation predicts that the equilibrium bond lengths of O-O and Se-O were found to be 1.45 Å and 2.00 Å, respectively as represented in Figure 12. The equilibrium structure of the OOH*_Nb-ZrSe$_2$ intermediate belongs to the *P1* (No.1) symmetry with equilibrium lattice constants a = 7.39 Å, b = 7.40 Å, and γ = 120.20° as reported in Table 3. The value of ΔG during the reaction O$_2$* + H$^+$ + e$^-$ → OOH* was found to be 0.13 eV. This positive value of the ΔG indicates the endothermic nature of the reaction. Thus, this step of ORR mechanism is unstable and readily proceed for the further reaction step(s). Furthermore, the value of ΔG is very close to zero, hence it can be considered ideal for ORR.

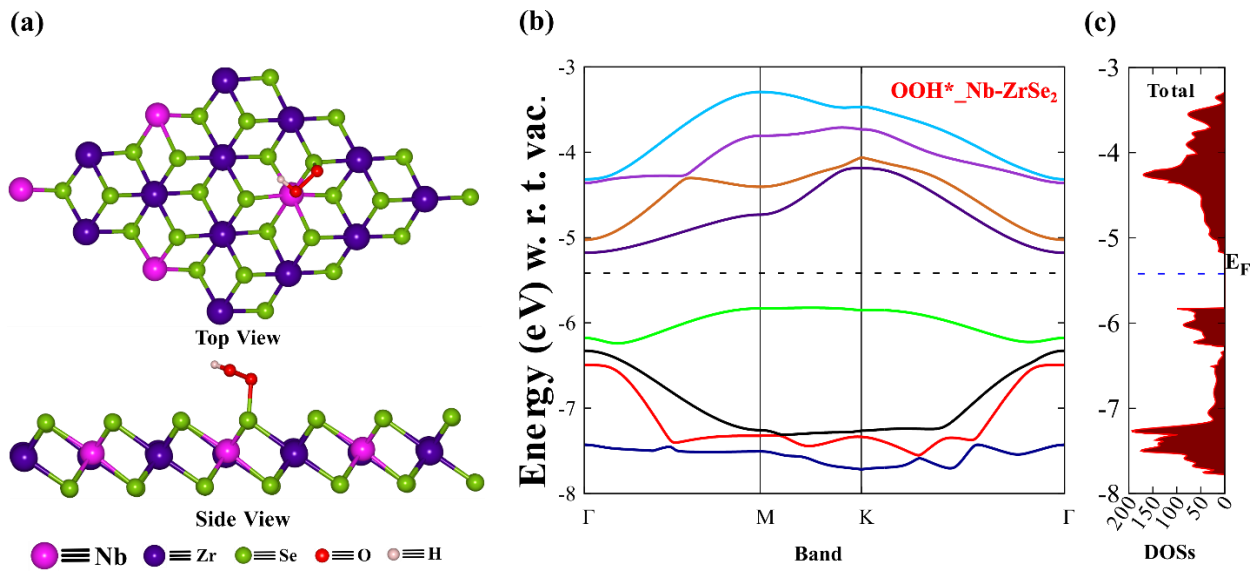

**Figure 7.** (a) Top and side view of equilibrium structure of OOH*_Nb-ZrSe$_2$, (b) band structure, and (c) total DOSs of OOH*_Nb-ZrSe$_2$.

**III$^{rd}$ step:** The next step of 4e$^-$ associative mechanism is the reduction of OOH* into O* through the process of hydrogenation of the reaction intermediate OOH*_Nb-ZrSe$_2$. We assume that a proton (H$^+$) coming from the anode side through proton exchange membrane and simultaneously, electron (e$^-$) flow through an external circuit, reacts with OOH* to form O* intermediate species. We have modeled this intermediate state by placing an oxygen atom at a distance 1.62 Å from the Se site which is close to equilibrium bond length of Se-O. The full optimization (lattice parameters and atomic coordinates) of the reaction intermediate O*_Nb-ZrSe$_2$ was performed and the equilibrium 2D layer structure (top and side view) of the O*_Nb-ZrSe$_2$ is represented in Figure 8a. Figure 8b-c represent the electronic band structure and DOSs profile of the O*_Nb-ZrSe$_2$ computed by DFT-D method. The band structures have been plotted along the high symmetry *Γ-M-K-Γ* k-path direction with respect to the vacuum followed by the previous computations. The $E_F$ of O*_Nb-ZrSe$_2$ system was found at -6.57 eV as shown by the dotted line in Figure 8b-c. From the band structure calculations of the O*_Nb-ZrSe$_2$ intermediate, it has been found that some of the energy bands cross the $E_F$, which indicates the metallic nature of the O*_Nb-ZrSe$_2$. The DOSs calculations of O*_Nb-ZrSe$_2$ show that there is electron density around the $E_F$ as shown in Figure 8c. So, both the band structure and DOSs profile shows the metallic nature of the O*_Nb-ZrSe$_2$, which supports the transportation of electrons involved in this reaction step. The equilibrium structure of the O*_Nb-ZrSe$_2$ intermediate has *P1* (layer group symmetry no. is 1) layer group symmetry with the optimized lattice parameters a = 7.37 Å, b = 7.41 Å, and γ = 120.10° as reported in Table 3. The equilibrium bond length of Se-O after the optimization was found to be 1.67 Å as represented in Figure 12. The value of ΔG in this reaction step (OOH* → O* + H$_2$O) was found to be -1.82 eV. The negative value of ΔG reveals a highly exothermic character of this reaction. This intermediate reaction step is thermodynamically stable and energetically favorable for ORR mechanism.

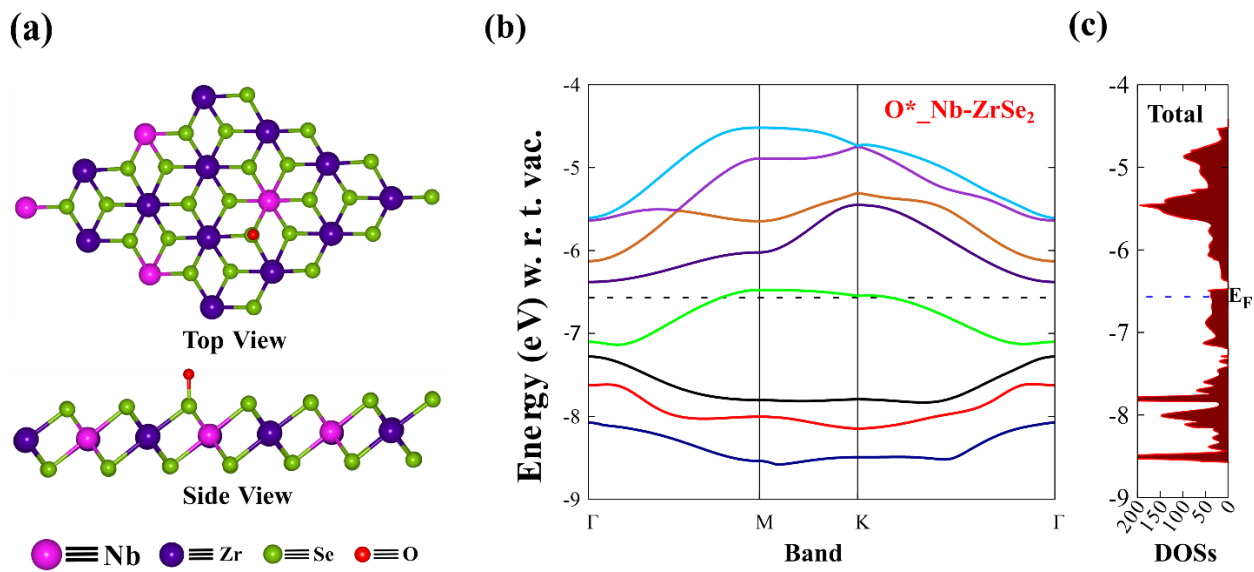

**Figure 8.** (a) Top and side view of equilibrium structure of O*_Nb-ZrSe$_2$, (b) band structure, and (c) total DOSs of O*_Nb-ZrSe$_2$.

**IV$^{th}$ step:** The next step of 4e$^-$ associative mechanism is the hydrogenation of the O*_Nb-ZrSe$_2$ intermediate by reacting with H$^+$ and e$^-$ coming from the anode side of the fuel cell. We assume that O* at the Se site capture one H$^+$ and e$^-$ to form OH* species. To model this intermediate state, we put an H atom at a distance 0.90 Å from the O atom, which is equilibrium bond distance of a free O-H molecule. The reaction intermediate OH*_Nb-ZrSe$_2$ was then optimized and the equilibrium structure of OH*_Nb-ZrSe$_2$ intermediate is illustrated in Figure 9a.

To examine the electronic behavior, we studied the electronic properties of the OH*_Nb-ZrSe$_2$

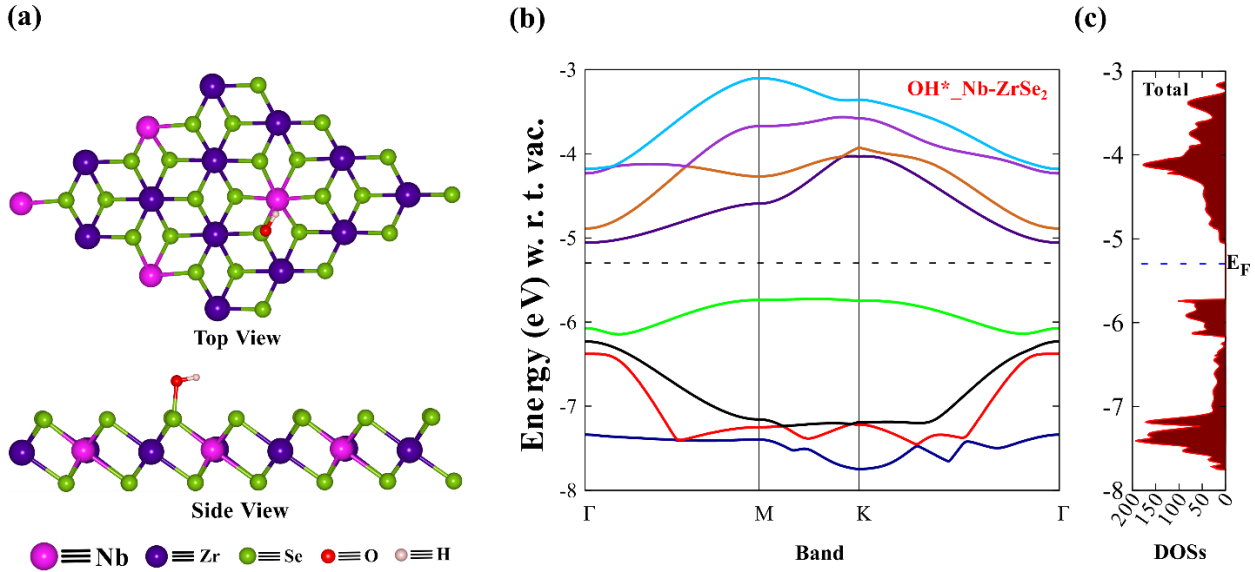

**Figure 9.** (a) Top and side view of equilibrium structure of OH*_Nb-ZrSe$_2$, (b) band structure, and (c) total DOSs of OH*_Nb-ZrSe$_2$.

intermediate. We have drawn the electronic band structure of the OH*_Nb-ZrSe$_2$ intermediate along the high symmetry k-path direction ***Γ-M-K-Γ*** in the first Brillouin zone with respect to the vacuum, which is consistent with all the intermediates with the pristine one. The E$_F$ of OH*_Nb-ZrSe$_2$ was found at -5.30 eV as depicted by the dotted line in Figure 9b-c. From the band structure calculations, we can see that the OH*_Nb-ZrSe$_2$ displays semiconducting nature with the indirect band gap of 0.66 eV as shown in Figure 9b. The conduction band minima and the valance band maxima of the OH*_Nb-ZrSe$_2$ are located at Γ and M, respectively, like previous intermediate step. To confirm this band gap, we drawn the DOSs profile corresponding to the band structures calculations, and the DOSs profile well supports the band gap of OH*_Nb-ZrSe$_2$ as shown in Figure 9c. The relaxed structure of the OH*_Nb-ZrSe$_2$ has *P1* (No.1) layer group symmetry with the equilibrium lattice constants a = 7.38 Å, b = 7.40 Å, and γ = 120.10° as reported in Table 3.

The equilibrium bond distance of Se-O and O-H were found to be 1.87 Å and 0.97 Å, respectively, as displayed in Figure 12. The value of ΔG for this reaction step (O* → OH*) was found to be -1.23 eV computed by the same DFT-D method. The negative value of ΔG indicates that this step is exothermic, and hence this process is thermodynamically and energetically favorable for the reduction of $O_2$. The value of ΔG of each reaction intermediates during associative mechanism has been reported in Table 4.

**Table 4.** The change in Gibbs free energy (eV) of the reaction steps during associative mechanism at the surface of the 2D monolayer Nb-ZrSe$_2$ material.

| Various reaction steps involved in associative mechanism | ΔG (eV) |
|---|---|
| [Nb-ZrSe$_2$+O$_2$ → O$_2$*_Nb-ZrSe$_2$] | -0.28 |
| [O$_2$*_Nb-ZrSe$_2$+H$^+$+e$^-$ → OOH*_Nb-ZrSe$_2$] | 0.13 |
| [OOH*_Nb-ZrSe$_2$+H$^+$+e$^-$ → O*_Nb-ZrSe$_2$+H$_2$O] | -1.82 |
| [O*_Nb-ZrSe$_2$+H$^+$+e$^-$ → OH*_Nb-ZrSe$_2$] | -1.23 |
| [OH*_Nb-ZrSe$_2$+H$^+$+e$^-$ → Nb-ZrSe$_2$+H$_2$O] | -1.76 |

**Reaction intermediates involved in dissociative mechanism:** The dissociative mechanism has the same initial step to initiate the ORR as associative mechanism i.e., $O_2$ adsorption on the Se-site of the 2D monolayer Nb-ZrSe$_2$. The next two steps will be different than associative mechanism, which we will discuss later in this manuscript. After that ORR process follows the same ORR steps. The dissociative mechanism involves the following distinct reaction steps:

**Dissociation of O$_2$* into 2O*:** After the adsorption of $O_2$ on the surface of the 2D monolayer Nb-ZrSe$_2$, the dissociation of $O_2$ into its constituents i.e., two O atoms at two different sites of the 2D

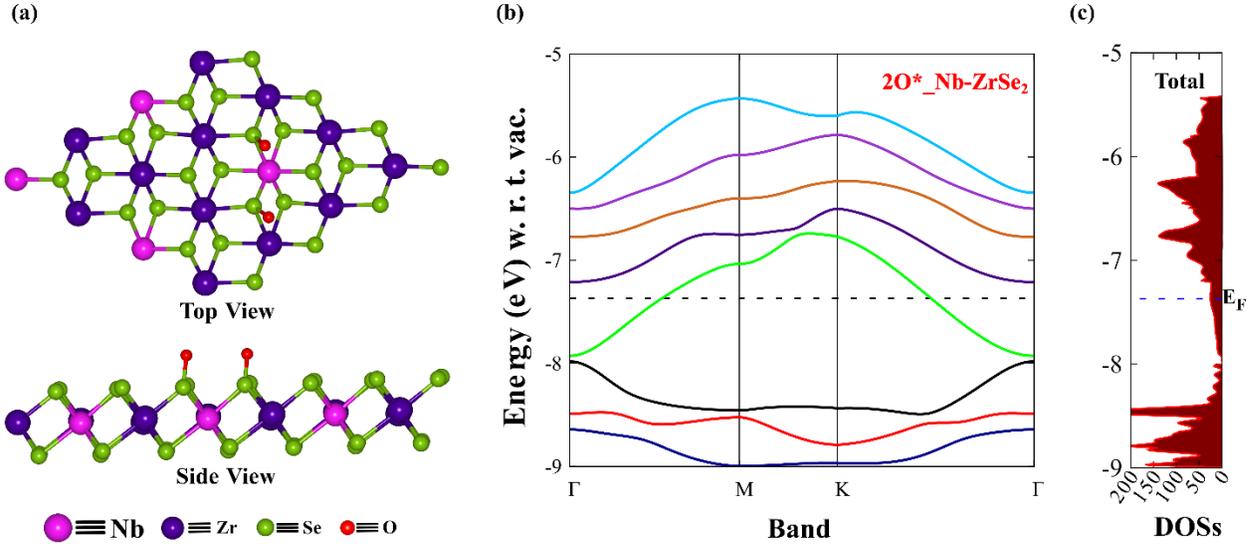

**Figure 10.** (a) Top and side view of equilibrium structure of 2O*_Nb-ZrSe$_2$, (b) band structure, and (c) total DOSs of 2O*_Nb-ZrSe$_2$.

Nb-ZrSe$_2$ occurs, and an intermediate has been formed during this dissociation reaction which is denoted by 2O*_Nb-ZrSe$_2$. To examine this reaction step, we put O atoms at a distance 1.61 Å from the Se-site of the 2D Nb-ZrSe$_2$. Then, we fully optimized the 2O*_Nb-ZrSe$_2$ structure and the relaxed equilibrium structure of 2O*_Nb-ZrSe$_2$ (top and side view) obtained by the DFT-D method is shown in Figure 10a. To examine the electronic behavior, we studied the electronic properties of the 2O*_Nb-ZrSe$_2$ system. We plotted the electronic band structure of the 2O*_Nb-ZrSe$_2$ intermediate along high symmetry k-path direction *Γ-M-K-Γ* in the first Brillouin zone. We plotted a total number of eight band with respect to the vacuum. From the band structure plot, we can see some of the electronic band crosses the E$_F$ indicating the metallic characteristic of the 2O*_Nb-ZrSe$_2$ as shown in Figure 10b. The E$_F$ of 2O*_Nb-ZrSe$_2$ is located at -7.37 eV as shown by dotted line in Figure 10b-c. To confirm conducting nature, we plotted the DOSs profile corresponds to the band structure plot. The DOSs profile shows a sufficient electron density around the E$_F$ as shown in Figure 10c. Thus, 2O*_Nb-ZrSe$_2$ has metallic nature which supports the electron transport during the reaction. The equilibrium structure of 2O*_Nb-ZrSe$_2$ has the *P1* (No.1) layer group symmetry with the equilibrium lattice constants a = 7.39 Å, b = 7.35 Å, and γ = 120.36° as reported in Table 3. After the optimization of this intermediate, the equilibrium bond length between Se and O atom was found to be 1.67 Å represented in Figure 12.

The value of ΔG for this reaction step (O$_2$* → 2O*) was calculated to be 1.90 eV as reported in Table 5. According to computed results, O$_2$* dissociation into 2O* is highly endothermic. Thus, we concluded that the dissociation of O$_2$* into 2O* is energetically unfavorable and thermodynamically unstable. Thus, the dissociative mechanism would be greatly suppressed by the associative mechanism. Thus, the ORR on the surface would favor the associative mechanism to reduce O$_2$ into H$_2$O.

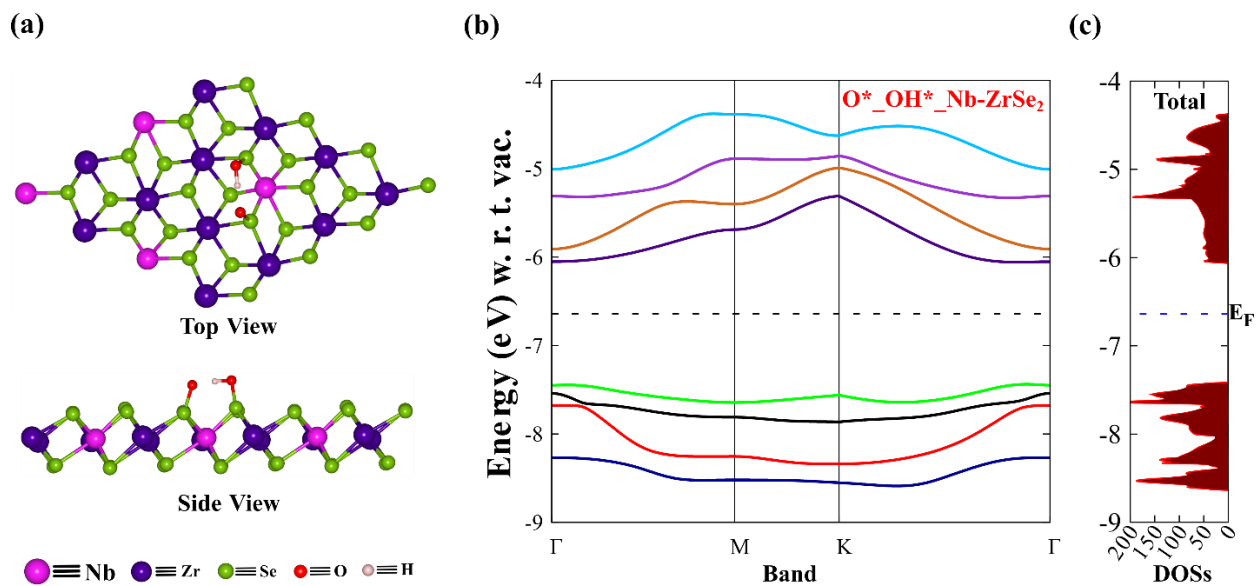

**Figure 11.** (a) Top and side view of equilibrium structure of O*_OH*_Nb-ZrSe$_2$, (b) band structure, and (c) total DOSs of O*_OH*_Nb-ZrSe$_2$.

**Protonation of 2O* to form O*_OH*:** Now the O$_2$ dissociation step is followed by the hydrogenation of 2O* by reacting with H$^+$ and e$^-$ coming from the anode side of the fuel cell. To examine this step, we have added a hydrogen atom at a distance 0.90 Å from one of the O* atoms. Thus, the intermediate state O*_OH* has been formed and the equilibrium structure was obtained by using the DFT-D method. The equilibrium structure (top and side view) of the O*_OH*_Nb-ZrSe$_2$ is depicted in Figure 11a.

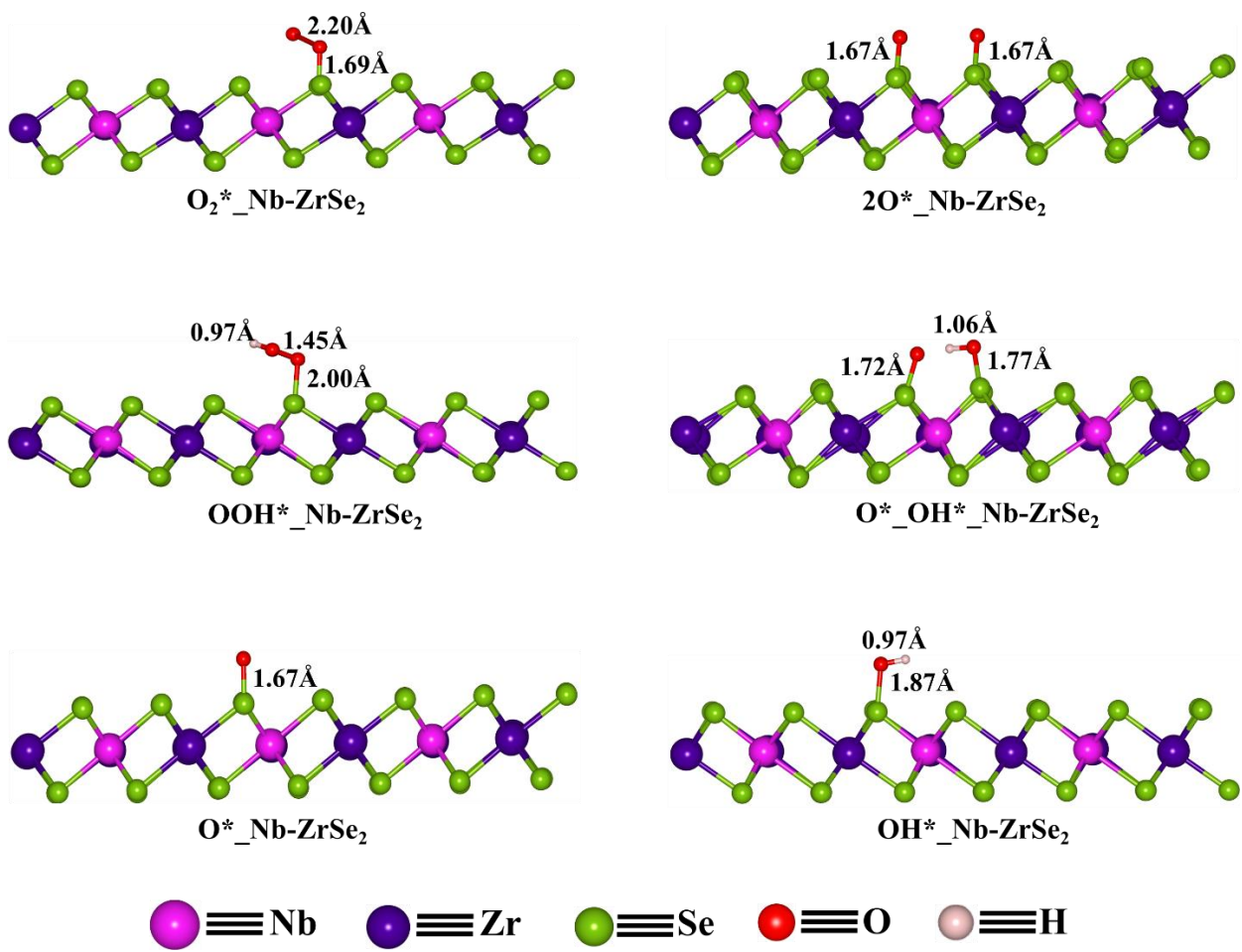

**Figure 12.** Structural properties of the various ORR intermediate structures.

The electronic properties calculations have been accomplished by using the same DFT-D method. We have calculated the electronic band structure along the high symmetric k-path direction *Γ-M-K-Γ* in the first Brillouin zone followed by the previous calculation. We can see that the electronic band structure of O*_OH*_Nb-ZrSe$_2$ intermediate has a direct band gap of 1.39 eV at the Γ point as shown in Figure 11b. Thus, the intermediate O*_OH*_Nb-ZrSe$_2$ shows a semiconducting nature. The E$_F$ is located at -6.64 eV as shown by the dotted line in Figure 11b-c. To confirm the electronic band gap of the O*_OH*_Nb-ZrSe$_2$, we also computed the DOSs profile corresponds to this band structures. The DOSs profile well supports the band gap of the O*_OH*_Nb-ZrSe$_2$ and the semiconducting nature of the O*_OH*_Nb-ZrSe$_2$ as shown in Figure 11c. The optimized structure has *P1* (No.1) layer group symmetry with equilibrium lattice constants a = 7.39 Å, b = 7.43 Å, and γ = 120.13° as reported in Table 3. The equilibrium Se-O, O-H, and Se-OH bond lengths have been found to be 1.72 Å, 1.06 Å, and 1.77 Å respectively, as represented in Figure 12. The value of ΔG for this reaction (2O* → O*_OH*) step was found to be -2.26 eV as reported in Table 5, which indicates this step of dissociative mechanism is exothermic and thermodynamically favorable.

**Table 5.** The change in Gibbs free energy of reaction steps in dissociative mechanism at the surface of the 2D monolayer Nb-ZrSe$_2$ material.

| Various reaction steps involved in dissociative mechanism | ΔG (eV) |
|---|---|
| [Nb-ZrSe$_2$+O$_2$ → O$_2$*_Nb-ZrSe$_2$] | -0.28 |
| [O$_2$*_Nb-ZrSe$_2$ → 2O*_Nb-ZrSe$_2$] | 1.99 |
| [2O*_Nb-ZrSe$_2$+H$^+$+e$^-$ → O*_OH*_Nb-ZrSe$_2$] | -2.26 |
| [O*_OH*_Nb-ZrSe$_2$+H$^+$+e$^-$ → O*_Nb-ZrSe$_2$+H$_2$O] | -1.40 |
| [O*_Nb-ZrSe$_2$+H$^+$+e$^-$ → OH*_Nb-ZrSe$_2$] | -1.23 |
| [OH*_Nb-ZrSe$_2$+H$^+$+e$^-$ → Nb-ZrSe$_2$+H$_2$O] | -1.76 |

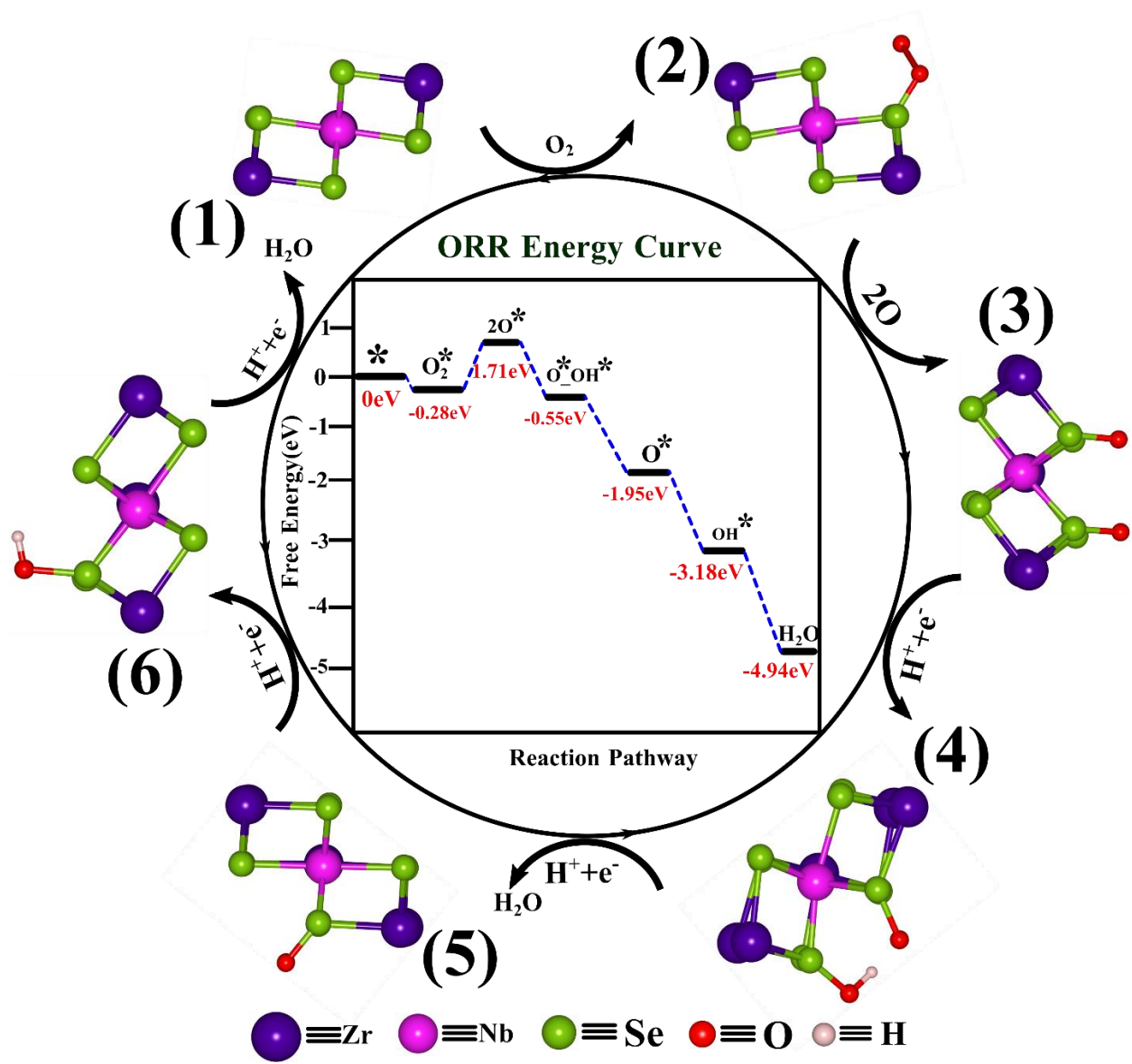

**Figure 13.** Free energy diagram of dissociative ORR mechanism at the surface of the 2D monolayer Nb-ZrSe$_2$ material.

After the protonation of 2O*, the dissociative mechanism follows the same intermediate states as the associative mechanism, which has been discussed earlier in this manuscript. The values of ΔG during the dissociative path have been reported in Table 5.

To summarize the catalytic activity of 2D Nb-ZrSe$_2$ towards ORR, we plotted and analyzed the ΔG curve (potential energy curve) for both associative and dissociative mechanism based on the DFT-D calculations as shown in Figure 13-14. We constructed these diagrams by considering the 2D Nb-ZrSe$_2$ as a reference geometry i.e., 2D Nb-ZrSe$_2$ considered corresponding to energy of 0 eV for both associative and dissociative mechanism of ORR. At a standard condition (i.e., at a temperature of 298.15 K with pressure 1 atm) the total value of ΔG for full reaction $O_2 + 2H_2 \rightarrow 2H_2O$ should be -4.92 eV. From our Gibbs energy landscape, the total value of ΔG was found to be -4.94 eV and -4.96 eV for dissociative and associative mechanism respectively. So, from our DFT-D calculations we can see that total value of ΔG for ORR mechanism are well consistent with the experimental value. In dissociative path, the free energy landscape shows that all the reaction steps are downhill except $O_2^* \rightarrow 2O^*$ as shown in Figure 13. For associative path, all the reaction steps are also downhill in the free energy landscape except $O_2^* \rightarrow OOH^*$ as shown in Figure 14. But by comparing both the free energy landscape, the $O_2^* \rightarrow 2O^*$ step of dissociative mechanism has higher uphill than that of the associative mechanism. Thus, the 2O* step is not stable and thermodynamically feasible for ORR mechanism, which makes the further ORR steps through dissociative path difficult. Whereas the uphill in associative mechanism is very small which is kinetically surmountable. Furthermore, all values of ΔG are negative, indicating that the entire ORR process through associative mechanism is exothermic and thermodynamically favorable. Therefore, the associative 4e$^-$ pathway for the ORR on the surface of 2D Nb-ZrSe$_2$ is the most favorable path.

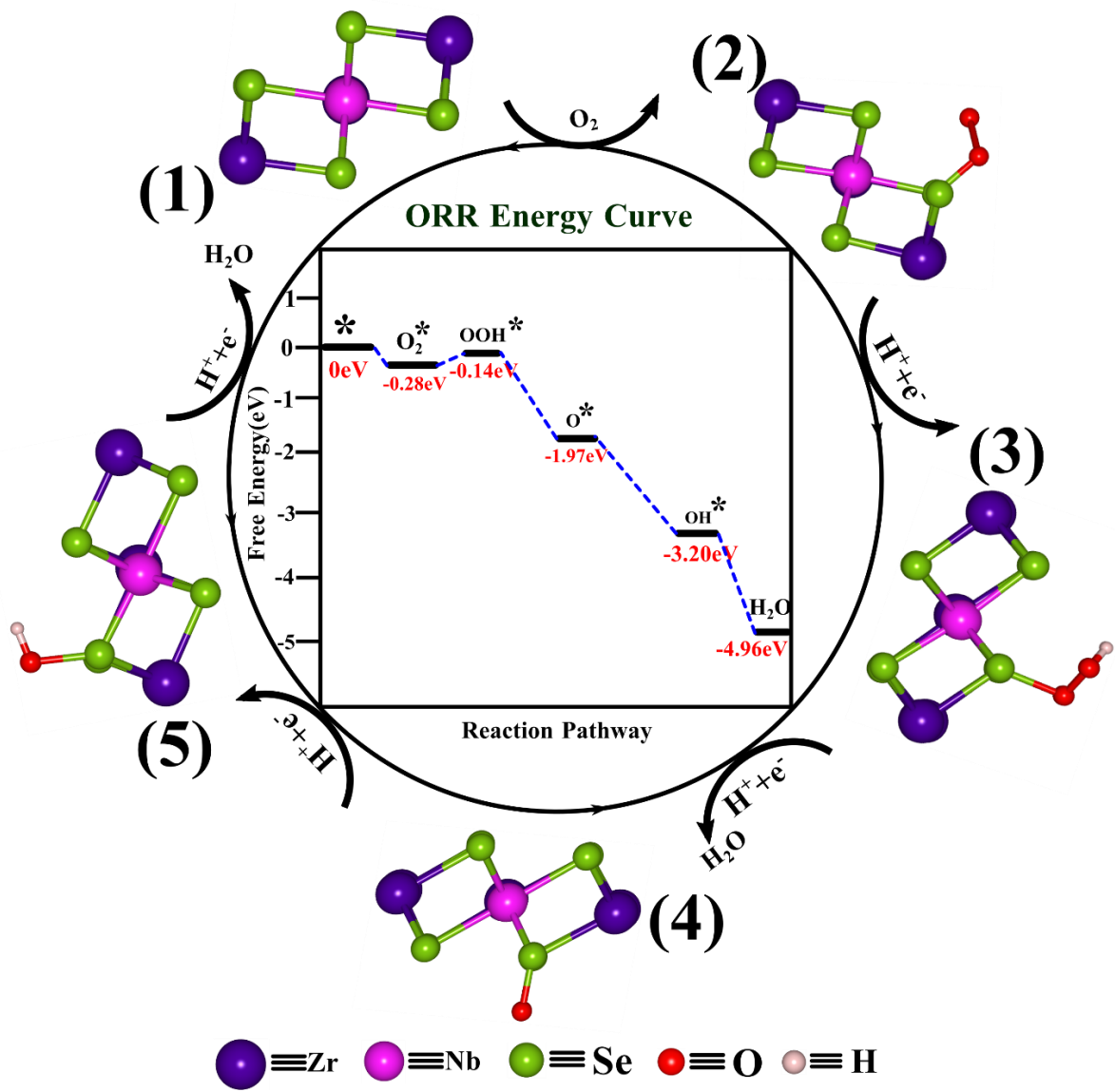

**Figure 14.** Free energy diagram of associative ORR mechanism at the surface of the 2D monolayer Nb-ZrSe$_2$ material.

**CONCLUSION**

In conclusion, we theoretically shed light on geometrical structure, electronic properties, and catalytic activity of 2D Nb-ZrSe$_2$ as well as detailed ORR mechanism on the surface of the 2D Nb-ZrSe$_2$ by using the DFT-D method. We investigated and examined both the possible path i.e., associative, and dissociative path of the ORR mechanism on the surface of the 2D Nb-ZrSe$_2$. The analysis of the electronic properties of pure monolayer ZrSe$_2$ demonstrated that a single layer of ZrSe$_2$ in a two-dimensional form exhibits semiconductor behavior, characterized by a band gap of 1.48 eV. To alter the electronic properties and catalytic activity of ZrSe$_2$, the substitutional doping of Nb has been done per 2x2 supercell of the ZrSe$_2$ monolayer. The 2D Nb-ZrSe$_2$ has zero band gap, indicating the metallic nature of 2D Nb-ZrSe$_2$. Conductivity of 2D Nb-ZrSe$_2$ plays a crucial role in facilitating the movement of electrons in the mechanism of ORR and 2D Nb-ZrSe$_2$ could act as an excellent catalyst towards ORR. To analyze the whole reaction process, ORR intermediate structures involved in the associative and dissociative path were constructed and optimized by using the same DFT-D method. To examine the catalytic activity, we computed the adsorption energy of ORR intermediates involved in both associative and dissociative mechanism. The dissociative mechanism involves the following intermediates O$_2$* → 2O* → O*_OH* → O* + H$_2$O → OH* and associative mechanism has the following intermediates O$_2$* → OOH* → O* + H$_2$O → OH*. For dissociative mechanism the value of ΔE of 2O* intermediate was found to be 1.99 eV to the catalytic site on the surface of the 2D Nb-ZrSe$_2$, which indicates that poor binding of 2O*. For associative mechanism the value of ΔE of OOH* intermediate was found to be -0.18 eV, which indicates that optimum binding of OOH* to the catalytic site on the surface of the 2D Nb-ZrSe$_2$. Thus, dissociative path would be less favorable path to reduce the O$_2$ molecule to water as compared to that of the associative path. In addition, for associative mechanism the value of ΔE for all intermediates is negative. The negative value of ΔE indicates that all the intermediates have good binding with Se site, thus energetically favorable path. The value of ΔG for all the reaction intermediates has also been computed. The values of ΔG also suggests that the associative path would be preferred over the dissociative mechanism. Thus, 2D Nb-ZrSe$_2$ can be used as an excellent catalyst for the ORR mechanism.

**Data and Software Availability.** This work was carried out with the following version of the programs. All the structures were modeled by using the VESTA software. (https://jp-minerals.org/vesta/en/) All the calculations were carried out by using the *ab-initio* based CRYSTAL17 suite code. (https://www.crystal.unito.it/) GNUPLOT and Inkscape 0.92 were used to plot the electronic band structures, total DOS, and PES. (http://www.gnuplot.info/ and https://inkscape.org/). Python scripts were used to generate data plots. (https://www.python.org/) The relevant data has been reported in Table 1, 2, 3, 4, and 5 in the main manuscript. All the equilibrium structures with their crystallographic information files (.cif) involved in the subject reaction have been provided in the Supporting Information. The data for this study can be obtained at https://pubs.acs.org/.

**Supporting Information**

The Supporting Information is available free of charge on the ACS Publications website. All the equilibrium structures involved in the subject reaction have been provided in the Supporting Information.

**Conflicts of interest**

There are no conflicts of interests to declare.

**AUTHOR INFORMATION**

**Corresponding Author: Dr Srimanta Pakhira**

Email: spakhira@iiti.ac.in or spakhirafsu@gmail.com

ORCID: 0000-0002-2488-300X34

**Corresponding Author**

**Dr. Srimanta Pakhira** − *Theoretical Condensed Matter Physics and Advanced Computational Materials Science Laboratory, Department of Physics, Indian Institute of Technology Indore (IIT Indore), Simrol, Khandwa Road, Indore-453552, Madhya Pradesh, India;*

*Theoretical Condensed Matter Physics and Advanced Computational Materials Science Laboratory, Centre for Advanced Electronics (CAE), Indian Institute of Technology Indore (IIT Indore), Simrol, Khandwa Road, Indore-453552, Madhya Pradesh, India;*

ORCID: orcid.org/0000-0002-2488-300X;

Email: spakhira@iiti.ac.in or spakhirafsu@gmail.com

**Authors**

**Mr. Ashok Singh** − *Theoretical Condensed Matter Physics and Advanced Computational Materials Science Laboratory, Department of Physics, Indian Institute of Technology Indore (IIT Indore), Simrol, Khandwa Road, Indore-453552, Madhya Pradesh, India.*

ORCID: orcid.org/0000-0002-2852-344X

**Acknowledgment**

We acknowledge the funding and technical support from the Science and Engineering Research Board-Department of Science and Technology (SERB-DST), Govt. of India under Grant No. CRG/2021/000572 and ECR/2018/000255. This research work is financially supported by the SERB-DST, Govt. of India under Grant No. CRG/2021/000572, ECR/2018/000255 and SB/S2/RJN-067/2017, SERB-DST, Govt. of India. Dr. Srimanta Pakhira acknowledges the Science and Engineering Research Board, Department of Science and Technology (SERB-DST), Government of India, for providing his highly prestigious Ramanujan Faculty Fellowship under the scheme no. SB/S2/RJN-067/2017, and for his Early Career Research Award (ECRA) under the grant No. ECR/2018/000255. Dr. Pakhira thanks to the SERB for providing the Core Research Grant (CRG), SERB-DST, Govt. of India under the scheme number CRG/2021/000572. Mr.

Ashok Singh thanks University Grants Commission (UGC), Govt. of India for availing his doctoral fellowship under the Ref. No: 1468/(CSIR-UGC NET JUNE 2019). The author would like to acknowledge the SERB-DST for providing the computing cluster and programs and IIT Indore for providing the basic infrastructure to conduct this research work. We acknowledge the National Supercomputing Mission (NSM) for providing computing resources of 'PARAM Brahma' at IISER Pune, which is implemented by C-DAC and supported by the Ministry of Electronics and Information Technology (MeitY) and Department of Science and Technology (DST), Government of India.

**Author Contributions:**

Dr. Pakhira developed the complete idea of this current research work, and Dr. Pakhira and Mr. Ashok Singh computationally studied the electronic structures and properties of the 2D monolayer Nb-ZrSe$_2$. Dr. Pakhira and Mr. Ashok Singh explored the whole reaction pathways, intermediates, and reaction barriers. Dr. Pakhira and Mr. Ashok Singh explained the ORR mechanism by the DFT calculations. Quantum calculations and theoretical models were designed and performed by Dr. Pakhira and Mr. Ashok Singh. Dr. Pakhira and Mr. Ashok Singh elucidated and analyzed the computed results and ORR mechanism. Dr. Pakhira and Mr. Ashok Singh wrote the whole manuscript and prepared all the Tables and Figures in the manuscript.

**GRAPHICAL ABSTRACT**

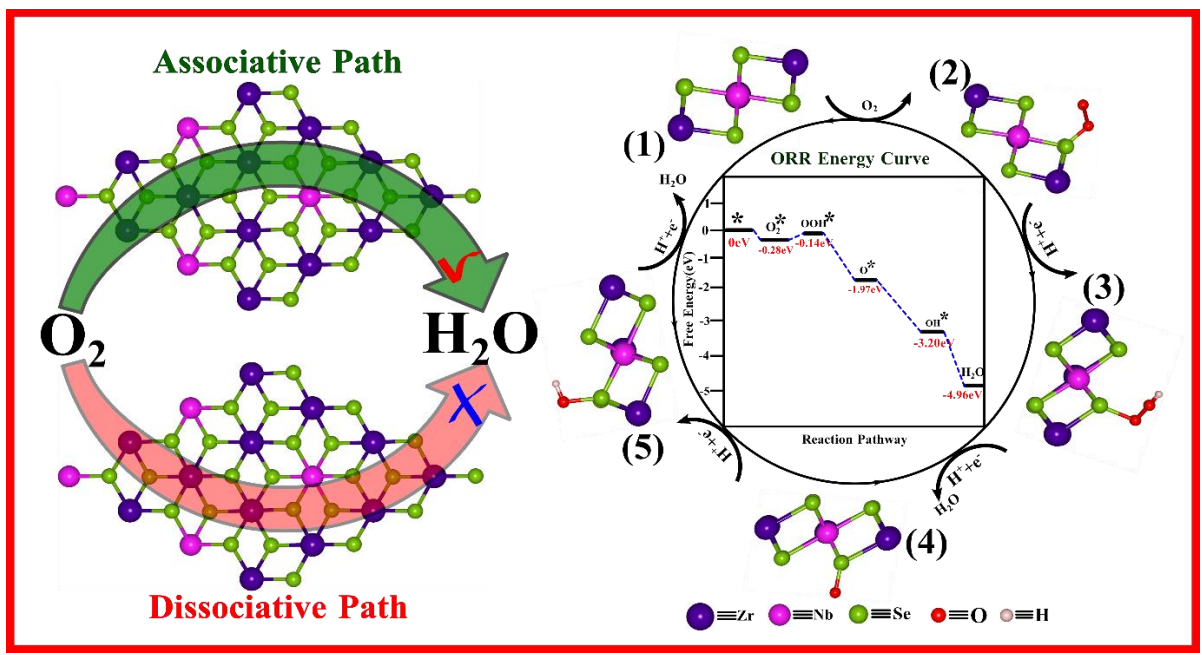